\documentstyle[prl,aps,epsf,twocolumn]{revtex}

\def\cM{{\cal M}}
\def\cZ{{\cal Z}}
\def\cP{{\cal P}}
\def\G{\Gamma}
\def\ba{\begin{array}}
\def\ea{\end{array}}
\def\be{\begin{equation}}
\def\ee{\end{equation}}
\def\bea{\begin{eqnarray}}
\def\eea{\end{eqnarray}}
\def\6{\partial}
\def\5{\bar }
\def\7{\tilde }
\def\8{\hat }
\def\a{\alpha}
\def\b{\beta}
\def\la{\lambda}
\def\ep{\epsilon}
\def\nn{\nonumber}

\def\state{\vert\alpha>}

\font\mybb=msbm10 at 10pt
\def\bb#1{\hbox{\mybb#1}}

\def\bR {\bb{R}}
\def\bI {\bb{I}}

\begin{document}

\title{
BPS States and Automorphisms
}

\author{Jordi Molins and
Joan Sim\'on $^\dag$}

\address{
Departament ECM,
Facultat de F\'{\i}sica,
Universitat de Barcelona and 
Institut de F\'{\i}sica d'Altes Energies,
Diagonal 647,
E-08028 Barcelona, Spain.
E-mail: molins@ecm.ub.es, jsimon@ecm.ub.es.\\
$^\dag$ Address after October 2, 2000 : Weizmann Institute,
Rehovot, Israel. \\[1.5ex]
\begin{minipage}{14cm}\rm\quad
The purpose of the present paper is twofold. In the first part,
we provide an algebraic characterization of several families of $\nu=
\frac{1}{2^n}$ $n\leq 5$ BPS states in M theory, at threshold and 
non-threshold, by an analysis of the BPS bound derived from the 
${\cal N}=1$ $D=11$ SuperPoincar\'e algebra. We determine their BPS masses 
and their supersymmetry projection conditions, explicitly. In the second part,
we develop an algebraic formulation to study the way BPS states transform
under $GL(32,\bR)$ transformations, the group of automorphisms of the
corresponding SuperPoincar\'e algebra. We prove that all $\nu=\frac{1}{2}$
non-threshold bound states are $SO(32)$ related with $\nu=\frac{1}{2}$ BPS 
states at threshold having the same mass. We provide further examples of 
this phenomena for less supersymmetric $\nu=\frac{1}{4},\frac{1}{8}$ 
non-threshold bound states.
\\[1ex]
PACS numbers: 11.10.Kk, 11.25.-w\\
Keywords: M-algebra, BPS states, automorphisms
\end{minipage}
}

\maketitle

%%%%%%%%%%%%%% body of paper %%%%%%%%%%%%%%%%%%%%%%%%%

\section{Introduction}

The importance of BPS states in string theory has largely been emphasized
in the literature. Not only they do provide a non-trivial check for the
conjectured web of dualities in string theory \cite{dual} but they are also
relevant for the microscopic computation of black hole entropies \cite{bhen}
and for studying non-perturbative phenomena in supersymmetric gauge field 
theories. It is therefore important to know their spectrum. One of the
purposes of the present paper is to go in this direction.

There are several ways to study the existence of BPS states in string theory.
Among them, one can look for classical supergravity solutions preserving
some amount of supersymmetry. This has been studied extensively, giving rise to
a large list of solutions involving intersections of branes and also
non-threshold bound states \cite{examp,eduardo}. One of the conclusions of all
these analysis was the derivation of some rules to construct BPS supergravity
solutions (harmonic superposition rule) and some techniques \cite{tech}
to generate new solutions from the previously known ones by T-duality
and M-reductions at angles and along boosted directions or through
electro-magnetic duality.

Besides world volume solitons, there exists certainly a third possibility,
which is entirely algebraic \cite{paulalgebra}, and which will be the one
followed in this paper. It consists in the resolution of the eigenvalue
problem associated with the saturation of the BPS bound derived from the
${\cal N}=1$ $D=11$ SuperPoincar\'e algebra. This approach characterizes
the Clifford valued BPS states $\state$ by its mass $\cM$ and the amount
of supersymmetry preserved $(\nu)$, which will generically be determined by 
some set of mutually commuting operators $\{\cP_i\}$ such that 
$\cP_i\state=\state$ $\forall i$. Both depend on the charges $\cZ$ carried
by $\state$. The main features of this eigenvalue problem are explained
in section II. Given two single $\nu=\frac{1}{2}$ BPS branes, their associated
supersymmetry projection operators, either commute or anticommute, giving rise
to an intersection of branes (BPS state at threshold) or a truly bound
state (non-threshold bound state) preserving $\nu=\frac{1}{4}$ or 
$\nu=\frac{1}{2}$ respectively. We generalize this idea by considering BPS 
states built up from two commuting or anticommuting BPS subsystems 
$(\cM_1 , \nu_1)$ and $(\cM_2 , \nu_2)$. In section III, we give the mass 
$\cM(\cM_1,\cM_2)$ and the amount of supersymmetry $\nu(\nu_1,\nu_2)$ 
preserved by the full state in both cases by proving two basic factoring 
theorems. Since their proof is constructive, we are also able to identify the 
corresponding set of projectors $\{\cP_i\}$ out of the ones $\{\cP_{1j}\}$ and
$\{\cP_{2j}\}$ characterizing the two subsystems. This approach is clearly
recursive, in the sense that any subsystem $(\cM_i,\nu_i)$ may be
decomposed into subsubsystems $(\cM_{ij},\nu_{ij})$ by application
of the previously mentioned theorems (factorizable states). It is in this way
that we identify and classify families of $\nu=\frac{1}{2^n}$ $n\leq 5$ BPS
states, both at threshold and non-threshold. Our analysis provides an 
intensive classification of the subset of BPS states in M-theory, and 
consequently in type IIA and type IIB theories, that we call factorizable
states. We give a large list of particular configurations satisfying our 
conditions, but certainly not an extensive one.

All the analysis done in section III is for an arbitrary value of the
central charges $\cZ$'s. In section IV, we first of all comment on some
of the general algebraic conditions that enhance supersymmetry through
fine tuning of the central charges $\cZ$'s \cite{gauntletthull}. Besides that,
we give an example for the easiest set of configurations not included in our
previous classification, the non-factorizable states.

The second purpose of this paper is to start analysing the way $\mbox{GL}(
32,\bR)$, the group of automorphisms of the ${\cal N}=1$ $D=11$ 
SuperPoincar\'e algebra \cite{west1,west2}, acts on such BPS states. We
first develop a useful algebraic formulation to check that indeed
$\mbox{GL}(32,\bR)$ is such a group and to compute the way central charges
$\cZ$ transform under such finite transformations. The subgroup of
$SO(32)$ transformations is identified with that preserving the mass
of BPS states. By studying some of the easiest $SO(32)$ transformations,
we prove that all non-threshold $\nu=\frac{1}{2}$ BPS states are $SO(32)$ 
related to $\nu=\frac{1}{2}$ BPS states at threshold of the same mass. 
We provide some particular examples of this phenomena, which is a 
generalization of the well-known fact that the truly bound state $\state$ 
characterized by $\G\state=\state$
\[
\G=\cos\alpha\,\G_{01} + \sin\alpha\,\G_{02}\,,
\]
associated with M-waves propagating in the 1-direction and 2-direction is 
related with an M-wave propagating in some intermediate direction 
$\vert\alpha^\prime>$, due to the existence of a rotation in the 12-plane 
$R_{12}$ belonging to the $SO(10)$ subgroup relating both states 
$(\state=R_{12}\vert\alpha^\prime>)$.

Even though we do not study the orbits of $\mbox{GL}(32,\bR)$, we do
provide evidence for the existence of more involved $SO(32)$ transformations
relating $\nu=\frac{1}{4}$ and $\nu=\frac{1}{8}$ non-threshold bound states 
to $\nu=\frac{1}{4}$ and $\nu=\frac{1}{8}$ BPS states at threshold having the 
same mass, thus partially generalizing the result on $\nu=\frac{1}{2}$ 
non-threshold bound states. We end up with some discussion concerning open 
questions related to world volume realization of automorphisms.

\section{Eigenvalue Problem}

The ${\cal N}=1$ $D=11$ SuperPoincar\'e algebra \cite{democracy,toine} is
basically described by
\begin{eqnarray}
\left\{ Q_\alpha, Q_\beta \right\} 
&=& (C\Gamma^{M})_{\alpha\beta} \cZ_M
+ \frac{1}{2} (C\Gamma_{MN})_{\alpha\beta} \cZ^{MN} \\
&&+ \frac{1}{5!} (C\Gamma_{MNPQR})_{\alpha\beta} \cZ^{MNPQR} \label{algsusy1}
\nonumber\\ 
\left[ Q_\alpha , \cZ^{M\dots} \right] &=& 0 \label{algsusy2}\ .
\end{eqnarray}
$Q_\alpha$ denotes the 32-component Majorana spinor
generating supersymmetry, whereas $\Gamma_{MN\dots}$ stands for
the antisymmetric product of $\Gamma_M$ matrices satisfying the Clifford 
algebra in eleven dimensions $\{\G_M , \G_N\}=2\eta_{MN}$, $M,N=0,1,\ldots
,9,\sharp$. The translation operators $P_M$ were denoted by $\cZ_M$, just by
notational convenience.

Since we are assuming that the above supersymmetry is valid at any energy,
the spectrum of M-theory should be organized into representations of the
SuperPoincar\'e algebra. We will be concerned with states preserving some
amount of supersymmetry, thus filling in short irreducible representations
of the latter algebra, the so called \emph{BPS states}. These can be entirely
characterized by purely algebraic techniques. In particular, given any 
M-theory state $\state$, the positivity of the matrix $<\alpha\vert\{Q_\a,
Q_\b\}\state$ implies a bound on the rest mass $\cM=\cZ^0$,
known as the Bogomol'ny bound. When the latter is saturated, there is a
linear combination of the supersymmetry generators annihilating the state.
This means that the symmetric matrix $\{Q_\a,Q_\b\}$ has at least one zero 
eigenvalue $\left(\det\,\{Q_\a,Q_\b\}=0\right)$. Thus, generically, the 
search for such BPS states is equivalent to the resolution of the eigenvalue 
problem
\be
\Gamma\state = \cM\state , 
\label{eigen}
\ee
where 
\[
\Gamma = \G_{0m}\cZ^m + \frac{1}{2}\G_0\G_{MN}\cZ^{MN} +
\frac{1}{5!}\G_0\G_{M_1\dots M_5}\cZ^{M_1\dots M_5}
\]
$m=1,\dots,9,\sharp$ and we have already used that the charged conjugation 
matrix may be chosen as $C=\G_0$.

The analysis of the corresponding eigenvalue problem was solved with full
generality for the ${\cal N}=1$ $D=4$ SuperPoincar\'e algebra in 
\cite{jeromepaul}, but the eleven dimensional problem is much more involved.
The first outstanding comment is that all antisymmetrized products of
Dirac matrices $(\G_i)$ appearing in the operator $\G=\sum_i \cZ^i\G_i$
satisfy
\be 
\Gamma_i^2=1, \  \mbox{tr} \ \Gamma_i = 0\,.
\label{cond0} 
\ee
We will call such matrices, \emph{single projectors}\footnote{Even though
they are not projectors in a strict sense.}.

Single projectors are related to $\nu=\frac{1}{2}$ BPS states corresponding
to single branes. This statement corresponds to the well-known fact that
given a single brane extending along certain directions of spacetime, or
equivalently, certain ``central charge'' $\cZ_1$, eq. (\ref{eigen})
becomes
\be
\cZ_1\G_1\state=\cM\state\,.
\label{single}
\ee 
Squaring eq. (\ref{single}), we derive that $\cM=\vert\cZ_1\vert$ and
$\G_1\state=\pm\state$, from which we infere that $\nu=\frac{1}{2}$, due to
(\ref{cond0}). In this way, we could rederive all $\nu=\frac{1}{2}$ BPS
states associated with single branes which are summarized in table 
\ref{table1}.

Actually, the mapping between single branes and single projectors is one to two
since given a single projector $\G_i$, there exists a second one $\7\G_i$
such that $\G_i\7\G_i=\bI$, since $\Gamma_0\Gamma_1 \cdots \Gamma_{\sharp} = 
\bI$.
This observation agrees perfectly with the known fact that the electrical
charges $\cZ^{0m}$ and $\cZ^{0m_1\dots m_4}$ correspond to M9-brane and
Mkk, respectively, since $\7\G_1=\G_{0m_1\dots m_9}$ and 
$\7\G_2=\G_{0m_1\dots m_6}$. From now on we will always be assuming that all
single projectors $\{\G_i\}$ appearing in our computations satisfy
$\G_i\G_j\neq \bI$ $\forall\, i,j$ $i\neq j$.

The second basic comment concerns the commutativity or anticommutativity
among two single projectors. That is, given any two single projectors
$\G_i\, , \G_j$, they either commute $[\G_i,\G_j]=0$ or anticommute
$\{\G_i,\G_j\}=0$. Let us consider the commuting case
\be
\G=\cZ_1\G_1 + \cZ_2\G_2 ,\ [\G_1,\G_2]=0\,.
\ee 
We obtain the new eigenvalue problem
\be
\G^\prime\state= 2\cZ_1\cZ_2\G_1\G_2\state=\cM^\prime\state\, ,
\label{eigen2}
\ee
where $\cM^\prime=\cM^2- (\cZ_1)^2 + (\cZ_2)^2$, by squaring eq. (\ref{eigen}).
Squaring eq. (\ref{eigen2}) fixes
\bea
\cM^\prime &=& \pm 2\cZ_1\cZ_2 \\
\G_1\G_2\state&=&\pm \state \, ,
\eea
that determines, upon substitution into the original eigenvalue problem,
the BPS mass
\be
\cM=\vert \cZ_1 \pm \cZ_2\vert
\ee
and the supersymmetry projection conditions
\be
\G_1\state=\G_2\state=\pm\state\,.
\ee
The above BPS state preserves $\nu=\frac{1}{4}$ since $[\G_1,\G_2]=0$
and $\mbox{tr}\left(\G_1\G_2\right)=0$, by hypothesis. Of course, we could
have derived the same conclusions by using an explicit representation
of the $\G$ matrices in a basis where both, $\G_1$ and $\G_2$ were diagonal.
We summarize all possible $\nu=\frac{1}{4}$ BPS configurations formed by
two commuting single branes in table \ref{table2}. We use the notation
$Mp\perp Mq(n)$, where $p$ and $q$ indicate the space dimensions along which
the branes are extended, whereas $n$ stands for the common space dimensions.

On the other hand, in the anticommuting case, when squaring eq. (\ref{eigen}),
we derive that the BPS mass is given by
\be
\cM=\sqrt{(\cZ_1)^2 + (\cZ_2)^2}\, ,
\ee
whereas the projection condition becomes
\be
\frac{\cZ_1\G_1 + \cZ_2\G_2}{\sqrt{(\cZ_1)^2 + (\cZ_2)^2}}\state=\pm\state\,.
\ee
It should be stressed that the coefficients appearing in the above condition
parametrize a circle of unit radius 
$\left(\cos\alpha=\cZ_1/\sqrt{(\cZ_1)^2 + (\cZ_2)^2}\right)$. The previous 
BPS state is $\nu=\frac{1}{2}$ due to the appearance
of a unique projection condition. It corresponds to a non-threshold
bound state, since $\cM\leq \cZ_1 + \cZ_2$. Alternatively, we could have
worked in a basis where
\bea
\G_1 &=& \bI_{16} \otimes \tau_3 \nonumber \\
\G_2 &=& \bI_{16} \otimes \tau_1 \,.
\eea
In this way, the condition of vanishing determinant becomes
\bea
& \left[\mbox{det}\left(\begin{array}{ccl}
\cZ_1 - \cM & \cZ_2 \\
\cZ_2 & -(\cZ_1 + \cM)
\end{array}\right)\right]^{16}= & \nonumber \\
& [\cM^2-((\cZ_1)^2 + (\cZ_2)^2)]^{16}=0 & \,,
\eea
from which we obtain the same conclusions as in the previous analysis.
As before, we summarize all possible non-threshold bound states
that one can construct out of two single branes in table \ref{table3}.

Up to now, we have solved eq. (\ref{eigen}) whenever $\G$ contains one or
two single projectors. One could continue in this direction classifying
the set of BPS states in terms of the number of single branes/projectors
$(N)$ involved in the eigenvalue problem (\ref{eigen}). Actually, once
$N$ is fixed, there are
$\left(\begin{array}{cl}
N \\
2
\end{array}\right)$ pairs of single projectors, giving rise to
$\left(\begin{array}{cl}
N \\
2
\end{array}\right)+1$ inequivalent configurations uncovering the set of all
inequivalent commutation relations among $N$ single projectors. It is
interesting to remark that one can map the problem of classifying the
set of initial $\G$ operators to the problem of classifying the inequivalent
graphs of $N$ points linked by $L$ lines \footnote{JS would like to thank
Ignasi Mundet for pointing out such a connection.}. One can draw a point
for every single projector involved in $\G$, and link any two of them
whenever the corresponding single projectors anticommute, leaving them
unlinked otherwise.

To solve the eigenvalue problem (\ref{eigen}) one can apply an algorithm
already used in \cite{pioline} and in the previous discussion. It consists
in acting on the left of eq. (\ref{eigen}) with $\G$, giving rise to
$\G^2\state = \cM^2\state$, which is equivalent to
\[
\left[\sum_{i=1}^N (\cZ_i)^2 + \sum_{i<j}^N\cZ_i\cZ_j\{\G_i , \G_j\}
\right]\state=\cM^2\state\,.
\]
Thus, whenever the set of $N$ single projectors $\{\G_i\}$ anticommute
$\{\G_i ,\G_j\}=0$ $\forall\, i,j$, the above equation already fixes
the eigenvalue to be 
\[
\cM^2 = \sum_{i=0}^N (\cZ_i)^2 \,,
\]
which when put it back into eq. (\ref{eigen}), gives rise to the
projection condition
\be
\sum_{i} \frac{\cZ_i}{\cM}\G_i \state = \pm \state \,,
\ee
which corresponds to a non-threshold bound state preserving $\nu=\frac{1}{2}$
build up of $N$ single projectors.

If there exist some pairs of commuting single projectors 
$\{\G_{\8\imath} ,\G_{\8\jmath}\}=2\G_{\8\imath}\G_{\8\jmath}$, one can map
equation $\G^2\state = \cM^2\state$ into a new eigenvalue problem
\be
\G^\prime\state=\cM^\prime\state\,,
\label{eigen3}
\ee
where
\[
\G^\prime=2\sum_{\8\imath<\8\jmath}^N \cZ_{\8\imath} 
\cZ_{\8\jmath} \G_{\8\imath}\G_{\8\jmath}\,,\]
and
\[
\cM^\prime = \cM^2 - \sum_{i=1}^N (\cZ_i)^2\,.
\]
Notice that $[\G,\G^\prime]=0$ since $\G^\prime = \G^2-(\cM^\prime -\cM^2)\bI$.
Given the new eigenvalue problem (\ref{eigen3}), one can proceed as for the
original one (\ref{eigen}). This defines an algorithm that will typically
finish whenever $\G^{(p)}$, the operator appearing in the p-th eigenvalue
problem, consists of a linear combination of anticommuting single projectors
\footnote{They are always linear combinations of single projectors due to
$\G_0\G_1\dots\G_9\G_\sharp=1$. For instance, from the definition of
$\G^\prime$, it is clear that $\G_{\8\imath}\G_{\8\jmath}$ are single 
projectors since $\left(\G_{\8\imath}\G_{\8\jmath}\right)^2=1$ and
$\mbox{tr}\left(\G_{\8\imath}\G_{\8\jmath}\right)=0$, by hypothesis.
Analogous arguments apply for $\G^{(p)}$ operators.}. 

Whenever the algorithm is solved, $\cM^{(p)}$ is fixed, and by construction,
the set of all $\cM^{(i)}$ $i=0,\dots ,p$ becomes fixed. Inserting these
eigenvalues back into the defining equations $\G^{(i)}\state=\cM^{(i)}\state$,
gives rise to a set of $(p+1)$ commuting projection conditions
\be
\cP^{(i)}\state=\state\, ,
\ee
determining the amount of preserved supersymmetry to be 
$\nu=\frac{1}{2^{p+1}}$.

Some important comments are in order at this point. First of all, that
$\nu=\frac{1}{2^{p+1}}$ relies on the fact that $\mbox{tr}\left(
\G^{(r)}\G^{(s)}\right)=0$ for $r\neq s$ and $r,s\leq p$. Furthermore, it
is assumed that no $\cP^{(i)}$ can be written as the product of other 
projectors. If this was the case, there would be an enhancement of
supersymmetry. The easiest example of this phenomena is the well-known
fact that one can add single branes for free in certain
configurations of intersecting branes. 

We would also like to stress that whenever $p>4$, the whole set of $32$
supercharges are already broken and the corresponding state does not belong
to any supersymmetry multiplet. For later convenience, it is worthwhile to
keep in mind that the above statements apply for arbitrary values of the
charges $\cZ_i$. There might be enhancement of supersymmetry for particular
values of the whole set of $\{\cZ_i\}$ under consideration.

\section{Factorizable States}

It is the aim of this section to describe some families of BPS states
solving the eigenvalue problem (\ref{eigen}). Before defining precisely
these families, we would like to present two \emph{factoring theorems}
that generalize previous results on two single brane configurations and allow
us to divide our original eigenvalue problem into simpler ones. Let us consider
the case in which the initial $\G$ operator decomposes according to
\be
\G :=\sum_a \cZ_a\Gamma_a + \sum_i \cZ_i\Gamma_i=:
\Gamma_{[a]} + \Gamma_{[i]}\,,
\label{subsystem}
\ee
where $a$ and $i$ label the first and second subsystem, respectively.
We will assume that both subsystems do have a solution of their associated
eigenvalue problems, i.e.
\bea
\Gamma_{[a]}\state&=&\cM_{[a]}\state \\
\Gamma_{[i]}\state&=&\cM_{[i]}\state\,.
\eea
Then, if $[\G_a , \G_i]=0$ $\forall\, a,i$, the eigenvalue problem
is solved by
\bea
\cM &=&\vert\cM_{[a]}\pm\cM_{[i]}\vert \\
\nu &=& \nu_{[a]}\cdot \nu_{[i]}\,,
\eea
whereas if $\{\G_a , \G_i\}=0$ $\forall\, a,i$, the solution is given by
\bea
\cM &=&\sqrt{\left(\cM_{[a]}\right)^2+\left(\cM_{[i]}\right)^2} \\
\nu &=& \frac{1}{2}\cdot \left(2 \nu_{[a]}\right)\cdot \left(2 \nu_{[i]}
\right)\,.
\eea
\emph{Proof}

\emph{(a)} If $[\G_a , \G_i]=0$ $\forall\, a,i$ $\Rightarrow [\Gamma_{[a]},
\Gamma_{[i]}]=0$. This means that we can simultaneously diagonalize 
$\Gamma_{[a]}$ and $\Gamma_{[i]}$ with eigenvalues $\pm \cM_{[a]}$,
$\pm \cM_{[i]}$, respectively. Thus $\cM=\vert \cM_{[a]} \pm \cM_{[i]} \vert$,
whereas the amount of preserved supersymmetry is 
$\nu=\nu_{[a]}\cdot \nu_{[i]}$. Actually, the set of projectors $\{\cP\}$
satisfying $\cP\state=+\state$\footnote{From now on, we will restrict
ourselves to the plus sign projection keeping in mind that both of them
$(\pm)$ are still possible}, is given by 
\be
\{\cP\}=\{\cP_{[a]},\cP_{[i]}\}\,.
\label{a}
\ee

\emph{(b)} If $\{\G_a , \G_i\}=0$ $\forall\, a,i$ $\Rightarrow \G^2=
\G_{[a]}^2 + \G_{[i]}^2$. But, $\G_{[a]}^2=\sum_a \left(\cZ_a\right)^2+
\G_{[a]}^\prime$ and 
$\G_{[i]}^2=\sum_i \left(\cZ_i\right)^2+ \G_{[i]}^\prime$. Thus the original
eigenvalue problem is mapped after the first step of the algorithm to
\be
\left(\G_{[a]}^\prime + \G_{[i]}^\prime\right)\state=\cM^\prime \state \,,
\nonumber
\ee
where $\cM^\prime = \cM^2 - \sum_a \left(\cZ_a\right)^2 -
\sum_i \left(\cZ_i\right)^2$. The important point is that
$[\G^\prime_a, \G^\prime_i]=0$ $\forall\, a,i$ since
$\G^\prime_a=2\cZ_b\cZ_c \G_b\G_c$ and $\G^\prime_i=2\cZ_k\cZ_l \G_k \G_l$.
We can then apply the previous result \emph{(a)} to fix
$\cM^\prime= \cM_{[a]}^\prime + \cM_{[i]}^\prime$, from which we derive that
\be
\cM=\sqrt{\cM_{[a]}^2 + \cM_{[i]}^2}\,,
\ee
since $\cM_{[a]}^2 = \cM_{[a]}^\prime + \sum_a \left(\cZ_a\right)^2$
and $\cM_{[i]}^2 = \cM_{[i]}^\prime + \sum_i \left(\cZ_i\right)^2$,
by hypothesis. Furthermore, $\nu^\prime=\nu_{[a]}^\prime\nu_{[i]}^\prime$
by \emph{(a)}. By construction, $\nu_{[a]}^\prime=2\nu_{[a]}$ and
$\nu_{[i]}^\prime=2\nu_{[i]}$. Finally, the amount of preserved supersymmetry
of the original problem is $\nu=\frac{1}{2}\nu^\prime$, since one must take
into account the original projection condition $\G\state=\cM\state$. We
conclude that $\nu = 2 \nu_{[a]}\cdot \nu_{[i]}$ as we claimed. Actually,
\be
\{\cP\}=\{\cP_0, \cP_{[a]}^\prime, \cP_{[i]}^\prime\}\,,
\label{b}
\ee
where $\cP_0$ stands for the original projector and $\cP_{[a]}^\prime$, 
$\cP_{[i]}^\prime$ stand for the projector conditions associated with 
$\G_{[a]}^\prime$, $\G_{[i]}^\prime$, respectively.

There is a nice geometrical picture emerging from our forementioned connection
to graph theory. Those BPS states satisfying the conditions for the first
factoring theorem correspond to disconnected graphs, whereas those associated
with the second theorem correspond to graphs containing subgraphs
$\left({\cal G}_i\right)$ such that all $N_i$ points belonging to ${\cal G}_i$
are linked with all $N_j$ points in ${\cal G}_j$ $\,i\neq j$. It would be
interesting to use known results from graph theory to classify the full
set of BPS states in M-theory, and viceversa, to use information from
branes into graph theory.

In the above proofs, we assumed the existence of a non-trivial solution
to $\G_{[a]}\state=\cM_{[a]}\state$ and $\G_{[i]}\state=\cM_{[i]}\state$,
so in this respect, they are completely general. In particular, we claim
that whenever $\G_{[a]}$ and $\G_{[i]}$ admit a similar decomposition
into commuting/anticommuting subsystems such that we can again apply the
factoring theorems, the original eigenvalue problem (\ref{eigen}) has a 
solution. By construction, all BPS states captured in this way will preserve
$\nu=\frac{1}{2^{p+1}}$ $p\leq 4$. We shall now describe, in more detail,
the forementioned BPS states, according to the value of $p$.

\subsection{$\nu=\frac{1}{2}$ BPS states}

It should be clear from our previous discussions that the most general
$\nu=\frac{1}{2}$ BPS state described by our hypothesis is given by a linear
combination of anticommuting single projectors. Actually, its mass is given
by
\[
\cM_{1/2}=\sqrt{\sum_{i=1}^N \left(\cZ_i\right)^2}\,,
\]
whereas the projection condition is
\[
\cP\state=\state \, , \, \cP = \frac{1}{\cM_{1/2}}\sum_{i=1}^N
\cZ_i\G_i \,.
\]
Equivalently, the coefficients in the linear combination 
$\cP=\sum_i a^i\G_i$ parametrize an $S^{N-1}$ sphere, so that they can always
be rewritten in terms of trigonometric functions.

Notice that these states depend on the number of single branes $(N)$ forming
them. For $N=1$, we recover the usual single branes (see table \ref{table1}), 
whereas for $N\geq 2$, we find families of non-threshold bound states, since 
their mass satisfies
\[
\cM_{1/2} \leq \sum_{i=1}^N \cM_{1/2}^{(i)}\,.
\]
 
All $N=2$ possibilities have already been summarized in table \ref{table3}. 
The latter
are the building blocks for higher $N$ non-threshold bound states. Generically,
to construct a $(\nu=\frac{1}{2},N+1)$ non-threshold bound state one must
just look for a single projector anticommuting with all $N$ single projectors
characterizing the original $(\nu=\frac{1}{2},N)$ bound state. Again, all this
information is already encoded in table \ref{table3}. Thus from 
$M2\perp M2(1)$, one can add a third M2-brane giving rise to two inequivalent 
arrays, either
\begin{equation}
\ba{cccccccccccl}
M2: &1&2&\_&\_&\_&\_&\_&\_&\_&\_ & \nn \\
M2: &\_&2&3&\_&\_&\_&\_&\_&\_&\_ & \nn \\
M2: &1&\_&3&\_&\_&\_&\_&\_&\_&\_ & 
\ea
\nonumber
\end{equation}
or
\begin{equation}
\ba{cccccccccccl}
M2: &1&2&\_&\_&\_&\_&\_&\_&\_&\_ & \nn \\
M2: &1&\_&3&\_&\_&\_&\_&\_&\_&\_ & \nn \\
M2: &1&\_&\_&4&\_&\_&\_&\_&\_&\_ & \,,
\ea
\label{gen}
\end{equation}
which from now on will be denoted by $2^3\{0,3,0\}$ and $2^3\{3,0,1\}$,
respectively, using the same notation as in \cite{eduardo}. Generically
$p_1^{r_1}\dots p_s^{r_s}\{n_1,\dots ,n_N\}$ will denote a configuration
of $N=r_1+ \dots + r_s$ branes of $p_1 ,\dots , p_s$ world space dimensions
such that the number of columns in the corresponding array with $i$ common
wordspace directions is $n_i$ $i\in\{1,\dots ,N\}$. Clearly, configuration
(\ref{gen}) can be extended to $2^N\{N,0,\dots ,1\}$ $1\leq N \leq 9$.

There is a large number of states belonging to this category, which we do not 
have the intention to classify extensively, the whole set being defined
in an intensive way by $\cP\state=\state$. Some particular examples are
given by $2^2 5^1\{2,2,1\}$, $2^1 5^2\{4,4,0\}$, $2^1 5^1 6^1\{6,2,1\}$,
$6^3\{4,2,3\}$, $6^2 9^1\{0,6,3\}$, $2^2 5^1 6^1\{6,0,3,0\}$,
$2^1 6^1 9^2\{2,2,6,1\}, \dots$

\subsection{$\nu=\frac{1}{4}$ BPS states}

There exist two possibilities to build up $\nu=\frac{1}{4}$ BPS states
from our factoring theorems. One can either apply the first one
giving rise to BPS masses
\be
\cM^{(1)}_{1/4} = \cM_{1/2} + \cM_{1/2} \,,
\label{m41}
\ee
or the second one, giving rise to
\be
\cM^{(2)}_{1/4} = \sqrt{\left(\cM_{1/2}\right)^2 + \left(\cM_{1/4}^{(1)}
\right)^2}\,.
\label{m42}
\ee
In the first case, the projection conditions are given by the union of the
subsystem projections (\ref{a})
\be
\cP_1\state=\cP_2\state=\state\,.
\label{m4a}
\ee 
In the second case, it can be seen by using the second factoring theorem
(\ref{b}) and applying the previously described algorithm that
\bea
\cP_2\cP_3\state &=& \state \nonumber \\
\cP_0\state &=& \state \,,
\label{m4b}
\eea
where $\cP_0 = \frac{1}{\cM_{1/4}^{(2)}}\left\{\cP_1\cM_{1/2} +
\cP_2\cM_{1/4}^{(1)}\right\}$.

Equations (\ref{m41}) and (\ref{m4a}) describe a threshold bound state
(intersection), whose components are the previously discussed $\nu=\frac{1}{2}$
BPS states. As such, they will be characterized by two integer positive numbers
$(N_1,N_2)$ describing the number of single branes in both subsystems. The
particular case $N_1=N_2=1$ corresponds to the standard intersection
of two branes (see table \ref{table2}). Now we see there are more general 
configurations involving non-threshold bound states in both subsystems. As 
before, all needed information is already contained in tables \ref{table2}
and \ref{table3}. For example, for $N_1=1$ and $N_2=2$ one can find, among 
many others
\bea
& 2^2 5^1\{3,3,0\} &  \nonumber \\
& 5^3\{6,3,1\} & \nonumber \\
& 2^3\{4,1,0\} & \nonumber \\
& 2^1 6^2\{6,1,2\} & \,, 
\label{412}
\eea
when $N_1=N_2=2$, 
\bea
& 2^2 5^2\{2,3,2,0\} &  \nonumber \\
& 5^4\{4,2,4,0\} & \nonumber \\
& 2^2 6^2\{6,0,2,1\} & \nonumber \\
& 2^4\{4,2,0,0\} & \,\dots 
\label{422}
\eea
It is straightforward to increase the values of $N_1 ,N_2$ by iteration of 
the latter process.

On the other hand, equations (\ref{m42}) and (\ref{m4b}) describe 
non-threshold bound states built from $\nu=\frac{1}{2}$ BPS states and the
above $\nu=\frac{1}{4}$ ones. As such, they will depend on three integer
numbers $(N_i)$. Let us comment on the easiest examples. When
$N_1=N_2=N_3=1$, take any configuration in table \ref{table2} and look for a 
third single brane whose associated single projector anticommutes with the 
latter two. For example, one can derive $2^2 5^1\{4,1,1\}$ from 
$M2\perp M5(1)$, $5^2 6^1\{2,7,0\}$ from $Mkk\perp M5(3)$ or 
$6^3\{4,4,2,0\}$ from $Mkk\perp Mkk(2)$.

When $N_1=1,\,N_2=2,\,N_3=1$, take any configuration from 
$\cM^{(1)}_{1/4}$ with $N_1=1$ $N_2=2$ and look
for a fourth single projector anticommuting with the latter three. In 
particular, from the examples written down in (\ref{412}), one can derive 
the existence of 
\bea
& 2^3 5^1\{3,1,2,0\} &  \nonumber \\
& 5^4\{5,1,3,1\} & \nonumber \\
& 2^4\{3,1,1,0\} & \nonumber \\
& 2^2 6^2\{7,1,1,1\} & \,, 
\label{4121}
\eea
respectively. When $N_1=N_2=2,\,N_3=1$ proceed as before, but starting from 
$\cM^{(1)}_{1/4}$ with $N_1=N_2=2$. One can
immediately find 
\bea
& 2^3 5^2\{2,3,0,2,0\} &  \nonumber \\
& 2^1 5^4\{4,2,2,2,0\} & \nonumber \\
& 2^2 5^1 6^2\{4,2,0,2,1\} & \nonumber \\
& 2^4 5^1\{4,3,1,0,0\} & \,\dots 
\label{4221}
\eea
from those quoted in (\ref{422}).

It should be clear that to increase $N^\prime_3=N_3 + j$ 
$(N_1,N_2 \,\mbox{fixed})$, one should look for $j$ single projectors
anticommuting with all projectors describing the $(N_1,N_2,N_3)$
configuration and among themselves. On the other hand, for a fixed $N_3$,
one should look for the subset of all $\cM^{(1)}_{1/4}$ configurations
whose $(N_1 + N_2)$ single projectors anticommute with all $N_3$
single projectors of the third subsystem. 

\subsection{$\nu=\frac{1}{8}$ BPS states}

Let us consider the different BPS states we can build from the first factoring
theorem. In this case, $\nu=\frac{1}{8}=\nu_1\cdot\nu_2 \Rightarrow
\nu_1=\frac{1}{2}\, , \nu_2=\frac{1}{4}$ (or viceversa), and we already
know there are two different ways to get $\nu_2=\frac{1}{4}$
BPS states. We are thus led to consider the BPS masses
\bea
\cM^{(1)}_{1/8} &=& \cM_{1/2} + \cM^{(1)}_{1/4} \label{m81} \\
\cM^{(2)}_{1/8} &=& \cM_{1/2} + \cM^{(2)}_{1/4}\,,
\label{m82}
\eea
with associated projection conditions
\bea
& \cP_1\state=\cP_2\state=\cP_3\state=\state & \label{p81} \\
& \cP_1\state=\cP_3\cP_4\state=\7\cP_2\state=\state \,,&
\label{p82}
\eea
corresponding to the joining of the subsystem projections, where
$\7\cP_2=\frac{1}{\cM^{(2)}_{1/4}}\left\{\cP_2\cM_{1/2} +
\cP_3\cM^{(1)}_{1/4}\right\}$.

Equations (\ref{m81}) and (\ref{p81}) describe $\nu=\frac{1}{8}$ BPS states
at threshold depending on three positive integers $(N_i)$. The particular
case $N_i=1$ $i=1,2,3$ is the well-known one involving triple intersections
of single branes. They are obtained from table \ref{table3} by looking for
a third single projector commuting with previous two. Examples involving
Mkk-monopoles or M9-branes are 
\bea
& 2^1 5^1 6^1\{6,2,1\}& \,, \nonumber \\
& 5^1 6^2\{6,1,3\} & \,, \nonumber \\
& 2^1 6^1 9^1\{4,5,1\}& \,, \nonumber \\
& \dots & \nonumber
\eea

The search of $N_1=1$ states is equivalent to looking for $\G_1$ single
projectors commuting with the projectors characterizing the $\cM^{(1)}_{1/4}$
subsystem $(N_2,N_3)$. In this way, by adding adequate single
branes one can derive the existence of
\bea
& 2^3 5^1 \{3,4,0,0\}&  \nonumber \\
& 2^1 5^3 \{4,3,2,0\} & \nonumber \\
& 2^4 \{6,1,0,0\} & \nonumber \\
& 2^1 5^1 6^2 \{2,5,1,1\} & \,, \nonumber
\eea
from the previously discussed $N_2=1$ $N_3=2$ $\cM^{(1)}_{1/4}$ states.
In exactly the same way, from the $N_2=N_3=2$ $\cM^{(1)}_{1/4}$ states
mentioned before we obtain,
\bea
& 2^3 5^2 \{3,2,3,0,0\} & \,, \nonumber \\
& 2^1 5^4 \{3,3,3,1,0\} & \,, \nonumber \\
& 2^3 6^2 \{4,2,2,1,0\} & \,, \nonumber \\
& 2^5 \{6,2,0,0,0\} & \,. \nonumber
\eea
To increase the value of $N_1$, keeping $N_2\,,N_3$ fixed, we must look for
$\G_i$ single projectors commuting with projectors characterizing 
$\cM^{(1)}_{1/4}$ and anticommuting with $\G_1$. We give some particular
examples for the first non-trivial example,  $N_i=2$ $i=1,2,3$
\bea
& 2^4 5^2 \{4,2,2,1,0,0\} & \,, \nonumber \\
& 2^2 5^4 \{2,4,3,0,1,0\} & \,, \nonumber \\
& 2^4 6^2 \{3,2,3,1,0,0\} & \,, \nonumber \\
& 2^6 \{6,3,0,0,0,0\} & \,. \nonumber
\eea

Equations (\ref{m82}) and (\ref{p82}) describe $\nu=\frac{1}{8}$
BPS states at threshold built from $\cM_{1/2}$ and $\cM^{(2)}_{1/4}$.
As such, they depend on four integer numbers. Generically, to construct
such states one should start from $\cM^{(2)}_{1/4}$ and look for a
projector $\cP_1$ commuting with $\cP_3\cP_4$ and $\7\cP_2$. We shall
provide some particular examples for the lower integer values. For instance,
when $N_i=1$ $\,i=1,2,3,4$ one may construct
\bea
& 2^3 5^1 \{4,2,1,0\} & \,, \nonumber \\
& 5^3 6^1 \{1,5,3,0\} & \,, \nonumber \\
& 2^1 6^3 \{4,2,4,0\} & \,, \nonumber \\
& 2^4 \{4,2,0,0\} & \,; \nonumber
\eea
when $N_i=1$ $\,i=1,2,3$ and $N_4=2$,
\bea
& 2^4 5^1 \{3,2,2,0,0\} & \,, \nonumber \\
& 2^5 \{5,1,1,0,0\} & \,, \nonumber \\
& 2^3 6^2 \{5,3,1,1,0\} & \,, \nonumber \\
& 5^5 \{3,3,1,2,1\} & \,; \nonumber
\eea
whereas for $N_i=1$ $N_j=2$ $\,i=1,2$ and $j=3,4$ we find, among others
\bea
& 2^4 5^2 \{3,2,1,2,0,0\} & \,, \nonumber \\
& 2^1 5^5 \{3,2,1,3,1,0\} & \,, \nonumber \\
& 2^5 5^1 \{4,4,1,0,0,0\} & \,, \nonumber \\
& 2^3 5^1 6^2 \{3,2,1,2,1,0\} & \,. \nonumber
\eea

On the other hand, if we intend to apply the second factoring
theorem, then $\nu=\frac{1}{8}=\frac{1}{2}2\nu_1\cdot 2\nu_2$ admits
as solutions either $\nu_1=\nu_2=\frac{1}{4}$ or $\nu_1=\frac{1}{2}$,
$\nu_2=\frac{1}{8}$. We are thus led to consider eight more families
of BPS states,
\bea
\cM^{(i+j+1)}_{1/8} &=& \sqrt{\left(\cM^{(i)}_{1/4}\right)^2 +
\left(\cM^{(j)}_{1/4}\right)^2} \, \mbox{i,j}=1,2 \label{m8i}\\
\cM^{(a+5)}_{1/8} &=& \sqrt{\cM^2_{1/2} + \left(\cM^{(a)}_{1/8}\right)^2}
\, \mbox{a}=1,\dots ,5 \label{m8a}
\eea
which correspond to non-threshold bound states preserving $\nu=\frac{1}{8}$.
Let us analyze the inequivalent configurations described by (\ref{m8i})
and (\ref{m8a}), in particular, their projection conditions.

$\cM^{(3)}_{1/8}$ describes a non-threshold bound state built from two
$\cM^{(1)}_{1/4}$ subsystems. It will thus be characterized by four integers.
As such, its projector conditions are given by
\bea
\cP_1\cP_2\state &=& \state \nonumber \\
\cP_3\cP_4 \state &=& \state \,,
\eea
while its non-threshold nature is due to
\be
\cP_0\state = \state\,,
\ee
where $\cP_0 = \frac{1}{\cM^{(3)}_{1/8}}\left\{\cP_1\cM^{(1)}_{1/4} +
\cP_3\cM^{\prime(1)}_{1/4}\right\}$. Particular examples with $N_i=1$ 
$\,i=1,2,3,4$ are
\bea
& 2^3 5^1 \{4,2,1,0\} & \,, \nonumber \\
& 2^2 5^1 6^1 \{4,2,1,1\} & \,, \nonumber \\
& 5^3 6^1 \{2,5,3,0\} & \,. 
\label{ea}
\eea
The general prescription would be to start from an $\cM^{(2)}_{1/4}$ 
configuration $(N_1,N_2,N_3)$ and to look for a single projector commuting
with $N_1$ single projectors and anticommuting with $N_2+N_3$ ones. In this 
way, from configurations (\ref{4121}) one may think of
\bea
& 2^3 5^2 \{3,3,1,1,0\} & \,, \nonumber \\
& 5^4 6^1 \{2,4,1,2,1\} & \,, \nonumber \\
& 2^4 5^1 \{5,1,2,0,0\} & \,, \nonumber \\
& 2^2 5^1 6^1 \{5,2,1,1,1\} & \,, \nonumber
\eea
respectively.

Analogously, $\cM^{(4)}_{1/8}$ describes a non-threshold bound state built from
two anticommuting $\cM^{(1)}_{1/4}$ and $\cM^{(2)}_{1/4}$ subsystems 
being characterized by five integers. That one
has $\cM^{(1)}_{1/4}$ is again described by
\be
\cP_1\cP_2\state = \state\,,
\ee
whereas $\cM^{(2)}_{1/4}$, which contains three subsystems $\{\cP_3,\cP_4,
\cP_5\}$, is described by
\be
\cP_4\cP_5\state = \state\,,
\ee
following the discussion on the second factoring theorem. The non-threshold
character of the state is inherited from
\be
\cP_0\state = \state\,,
\ee
where $\cP_0 =\frac{1}{\cM^{(4)}_{1/8}}\left\{\cP_1\cM^{(1)}_{1/4} +
\cP_3\cM_{1/2} + \cP_4\cM^{\prime (1)}_{1/4}\right\}$. The general prescription
to build up such states would be to start from an $\cM^{(3)}_{1/8}$ 
configuration $(N_1,\dots ,N_4)$ and to look for a single projector
anticommuting with all $N_1+\dots +N_4$ previous ones. To illustrate this
point, one can check that the following configurations can be derived from
(\ref{ea})
\bea
& 2^3 5^2 \{4,1,2,1,0\} & \,, \nonumber \\
& 2^3 5^1 6^1 \{3,3,1,0,1\} & \,, \nonumber \\
& 5^3 6^2 \{1,3,4,2,0\} & \,. 
\label{eb}
\eea

Similarly, BPS states $\cM^{(5)}_{1/8}$ depending on six integers are 
characterized by
\bea
\cP_2\cP_3\state &=& \state \nonumber \\
\cP_5\cP_6\state &=& \state \nonumber \\
\cP_0\state &=& \state 
\eea
where 
\[
\cP_0 = \frac{1}{\cM^{(5)}_{1/8}}\left\{\cP_1\cM_{1/2} + 
\cP_2\cM^{(1)}_{1/4} + \cP_4\cM^\prime_{1/2}
+ \cP_5\cM^{\prime(1)}_{1/4}\right\}\,.
\] 
Starting from $\cM^{(4)}_{1/8}$ configurations and searching for single
projectors anticommuting with the ones describing $\cM^{(4)}_{1/8}$ gives
rise to these BPS states. The following examples
\bea
& 2^3 5^3 \{5,0,1,2,1,0\} & \,, \nonumber \\
& 2^3 5^1 6^2 \{3,1,3,1,1,0\} & \,, \nonumber \\
& 5^3 6^3 \{1,1,4,2,2,0\} & \,, \nonumber
\eea
were generated from (\ref{eb}).

Concerning $\cM^{(a+5)}_{1/8}$, they are non-threshold bound states 
depending on $N_{a+5}=N_a + 1$ integers, built from
$\cM_{1/2}$ and $\cM^{(a)}_{1/8}$ subsystems. As such, the prescription
to generate these states is the same as the one explained for
$\cM^{(4)}_{1/8}$ and $\cM^{(5)}_{1/8}$. They are described by three 
projection conditions : two of them 
characterizing the $\cM^{(a)}_{1/8}$ subsystem and a third one
associated with its non-threshold nature.

\subsection{$\nu=\frac{1}{16}$, $\frac{1}{32}$ BPS states}

We would like to conclude with a general discussion concerning
$\nu=\frac{1}{16}$, $\frac{1}{32}$ BPS states extending previous techniques.
From the first factoring theorem, whenever $\nu=\frac{1}{16}=\nu_1\cdot\nu_2
\Rightarrow \nu_1=\frac{1}{2}\,,\nu_2=\frac{1}{8}$ and $\nu_1=\nu_2=
\frac{1}{4}$, whereas for $\nu=\frac{1}{32} \Rightarrow \nu_1=\frac{1}{2}
\,, \nu_2=\frac{1}{16}$ and $\nu_1=\frac{1}{4} \,, \nu_2=\frac{1}{32}$.
There will thus be BPS masses of the form
\bea
\cM^{(i)}_{1/16} &=& \cM_{1/2} + \cM^{(i)}_{1/8} \\
\cM^{(ij)}_{1/16} &=& \cM^{(i)}_{1/4} + \cM^{(j)}_{1/4} \\
\cM^{(i)}_{1/32} &=& \cM_{1/2} + \cM^{(i)}_{1/16} \\
\cM^{(ij)}_{1/32} &=& \cM^{(i)}_{1/4} + \cM^{(j)}_{1/8}\,,
\eea
whose projection conditions are given by the union of the subsystem
ones (see (\ref{a})).

If we apply the second factoring theorem, $\nu=2\nu_1\cdot\nu_2$.
For $\nu=\frac{1}{16} \Rightarrow \nu_1=\frac{1}{4} \nu_2=\frac{1}{8}$
or $\nu_1=\frac{1}{2} \nu_2=\frac{1}{16}$. On the other hand, when
$\nu=\frac{1}{32} \Rightarrow \nu_1=\frac{1}{4} \nu_2=\frac{1}{16}$,
$\nu_1=\nu_2=\frac{1}{8}$ or $\nu_1=\frac{1}{2} \nu_2=\frac{1}{32}$.
We summarize the table of BPS masses corresponding to the above
discussion as follows
\bea
\cM^{(ij)}_{1/16} &=& \sqrt{\left(\cM^{(i)}_{1/4}\right)^2+
\left(\cM^{(j)}_{1/8}\right)^2} \\
\cM^{(i)}_{1/16} &=& \sqrt{\left(\cM_{1/2}\right)^2+
\left(\cM^{(i)}_{1/16}\right)^2} \\
\cM^{(ij)}_{1/32} &=& \sqrt{\left(\cM^{(i)}_{1/4}\right)^2+
\left(\cM^{(j)}_{1/16}\right)^2} \\
\cM^{(ij)}_{1/32} &=& \sqrt{\left(\cM^{(i)}_{1/8}\right)^2+
\left(\cM^{(j)}_{1/8}\right)^2} \\
\cM^{(i)}_{1/32} &=& \sqrt{\left(\cM_{1/2}\right)^2+
\left(\cM^{(i)}_{1/32}\right)^2}\,.
\eea
Their projection conditions can be written by applying the same
methodology ussed for $\cM_{1/8}$ non-threshold BPS states.
Just to illustrate this general methodology, let us consider two particular
examples. To begin with, we shall analyze $\cM_{1/16}=
\sqrt{\left(\cM^{(1)}_{1/4}\right)^2+\left(\cM^{(7)}_{1/8}\right)^2}$. This
is a $\nu=\frac{1}{16}$ non-threshold BPS state. It consists on
$\cM^{(1)}_{1/4}$ and $\cM^{(7)}_{1/8}$ subsystems. The first subsystem
depends on two subsubsystems $\left(\cP_1\,,\cP_2\right)$ satisfying
\be
\cP_1\cP_2\state=\state\,,
\ee
whereas the second subsystem depends on five of them 
$\left(\cP_3\,,\cdots ,\cP_7\right)$ such that
\bea
\cP_4\7\cP_5 \state &=& \state \nonumber \\
\cP_6\cP_7\state &=& \state
\eea
where $\7\cP_5=\frac{1}{\cM^{(2)}_{1/4}}\left\{\cP_5\cM_{1/2} +
\cP_6\cM^{(1)}_{1/4}\right\}$. The final projection condition
describing the non-threshold nature of $\cM_{1/16}$ is given by
\be
\cP_0\state=\state\,,
\ee
where $\cP_0 = \frac{1}{\cM_{1/16}}\left\{\cP_1\cM^{(1)}_{1/4} +
\cP_3\cM_{1/2} + \cP_4\cM^{(2)}_{1/8}\right\}$\,. One brane configuration
satisfying such requirements is given by $2^2 5^3 6^2\{2,1,3,3,0,1,0\}$
corresponding to the array
\begin{equation}
\ba{cccccccccccl}
M5: &1&2&3&4&5&\_&\_&\_&\_&\_ & \nn \\
Mkk: &1&2&3&\_&\_&6&7&8&\_&\_ & \nn \\
M5: &\_&\_&\_&4&5&6&7&\_&9&\_ & \nn \\
M5: &1&\_&\_&4&5&6&\_&8&\_&\_ & \nn \\
M2: &1&2&\_&\_&\_&\_&\_&\_&\_&\_ & \nn \\
M2: &1&\_&3&\_&\_&\_&\_&\_&\_&\_ & \nn \\
Mkk: &1&\_&3&4&5&\_&\_&8&\_&\sharp & \,,
\ea
\end{equation}

Finally, consider the $\nu=\frac{1}{32}$ BPS state with mass $\cM_{1/32}=
\sqrt{\left(\cM^{(3)}_{1/8}\right)^2 + \left(\cM^{(5)}_{1/8}\right)^2}$.
This is a $\nu=\frac{1}{32}$ non-threshold BPS state. It consists on
$\cM^{(3)}_{1/8}$ and $\cM^{(5)}_{1/8}$ subsystems. The first one
depends on four subsystems $\left(\cP_1 ,\dots ,\cP_4\right)$
satisfying
\bea
\cP_1\cP_2\state&=&\state \nonumber \\
\cP_3\cP_4\state&=&\state
\eea
whereas the second subsystem depends on five of them $\left(\cP_5 ,\dots ,
\cP_{9}\right)$ such that
\bea
\cP_5\cP_6\state&=&\state \nonumber \\
\cP_{8}\cP_{9}\state&=&\state\,.
\eea
The non-threshold nature is given by
\bea
\cP_0\state &=&\state \nonumber \\
\cP_0 &=& \frac{1}{\cM_{1/32}}\left\{\cP_1\cM^{(1)}_{1/4} + 
\cP_3\cM^{\prime(1)}_{1/4} + \cP_5\cM^{\prime\prime}_{1/4} \right. \nonumber \\
& & \left. + \cP_7\cM_{1/2}+
\cP_8\cM^{\prime\prime\prime(1)}_{1/4} \right\}\,.
\eea
An example for this kind of non-threshold bound state is provided by
$2^6 5^2 9^1\{5,1,0,0,0,4,0,0,0\}$ or in terms of its array
\begin{equation}
\ba{cccccccccccl}
M2: &1&2&\_&\_&\_&\_&\_&\_&\_&\_ & \nn \\
M2: &\_&\_&3&4&\_&\_&\_&\_&\_&\_ & \nn \\
M2: &1&\_&3&\_&\_&\_&\_&\_&\_&\_ & \nn \\
M2: &\_&2&\_&4&\_&\_&\_&\_&\_&\_ & \nn \\
M2: &1&\_&\_&4&\_&\_&\_&\_&\_&\_ & \nn \\
M2: &\_&2&3&\_&\_&\_&\_&\_&\_&\_ & \nn \\
M5: &1&2&3&4&5&\_&\_&\_&\_&\_ & \nn \\
M9: &1&2&3&4&5&6&7&8&9&\_ & \nn \\
M5: &1&2&3&4&\_&\_&\_&\_&\_&\sharp & \,.
\ea
\end{equation}

\section{Further BPS states}

It should be clear that the set of BPS states described in the previous
section is just a subset of the full set of BPS states in M-theory. First
of all, the existence of BPS states corresponding to branes intersecting
at angles is already known. There has been a lot of work in this direction 
\cite{angles}, but the main idea behind their classification was based
on the resolution of eq. (\ref{eigen}), since one starts from
a set of parallel single branes, thus breaking $\nu=\frac{1}{2}$, and
rotate the projection condition $R\cP_i R^{-1}\state=\state$. In this way,
one retains the interpretation as an intersection of branes where the second
brane intersects the first one at angles. We will have nothing more to say 
about these configurations. Besides that, we would like to comment on the
possible existence \footnote{By possible, we mean not forbidden by pure
algebraic considerations.} of exotic branes (appearing by fine tuning the 
value of the central charges) and non-factorizable BPS states, that is, those 
not classified by the criteria introduced in section III.

\subsection{Exotic Branes}

In section II, we assumed the values of the charge operators $\{\cZ\}$
were arbitrary, but fixed. It was already pointed out in \cite{gauntletthull},
that for certain configurations, there exist very precise values of $\{\cZ\}$
giving rise to BPS configurations preserving \emph{exotic} fractions of
supersymmetry. In particular, they considered the M-theory configuration
\bea
\ba{cccccccccccl}
M5: &1&2&3&4&5&\_&\_&\_&\_&\_ & \nn \\
M5: &\_&\_&\_&\_&5&6&7&8&9&\_ & \nn \\
M2: &\_&\_&\_&\_&5&\_&\_&\_&\_&\sharp & \,.
\ea
\eea
The latter is described by three mutually commuting single projectors 
$(\G_{(1)}=\G_{012345}\, , \G_{(2)}=\G_{056789}\, ,\G_{(3)}=\G_{05\sharp})$,
where the third condition  $\G_{(3)}\state=\state$
can be derived from $\G_{(1)}\state=\G_{(2)}\state=\state$. Equivalently,
the M2-brane can be added for free.

Actually, any configuration consisting on three commuting single branes
such that one of them is for free will admit exotic fractions of
supersymmetry. The proof follows from the analysis done in
\cite{gauntletthull}. We can simultaneously diagonalize the three single
projectors. In particular,
\bea
\Gamma_1 &=& diag(1,1,-1,-1)\otimes \bI_8 \nn \\
\Gamma_2 &=& diag(1,-1,1,-1)\otimes \bI_8 \nn \\
\Gamma_3 &=& diag(1,-1,-1,1)\otimes \bI_8 \, ,
\label{dec}
\eea 
where we restricted ourselves to $\Gamma_3=\Gamma_1\Gamma_2$, without loss
of generality. In this case, the eigenvalue problem is 
equivalent to
\[
\{Q,Q\}=diag(\cM-\la_1,\cM-\la_2,\cM-\la_3,\cM-\la_4)\otimes \bI_8\, .
\]
The positivity of its left hand side implies that $\cM\geq \la_i$ 
$\forall\,i$, where
\bea
\la_1 &=& \cZ_1 + \cZ_2 + \cZ_3\nn \\
\la_2 &=& -\la_1 + 2\cZ_1 \nn \\
\la_3 &=& -\la_1+ 2\cZ_2 \nn \\
\la_4 &=& -\la_1+2\cZ_3 \, . 
\eea
It is clear that $\la_2$ is the biggest eigenvalue whenever
$0 \geq \cZ_1 \geq \cZ_2$ and $0 \geq \cZ_1 \geq \cZ_3$, so that the BPS bound
becomes 
\bea
\cM\geq -\la_1 + 2\cZ_1 \,.
\eea
When the latter is saturated, and for arbitrary values of $\cZ_1,\cZ_2,\cZ_3$
satisfying the latter restrictions, then
\bea
\{Q,Q\}&=&diag(-2(\cZ_2+\cZ_3),0,2(\cZ_1-\cZ_2),2(\cZ_1-\cZ_3))\nonumber \\
& & \otimes \bI_8
\eea
the corresponding BPS state preserves $1/4$ of susy.
Notice that whenever $\cZ_1=\cZ_2$ or $\cZ_1=\cZ_3$, the amount of susy becomes
enhanced to $1/2$, and the mass becomes $\cM=\vert \cZ_3\vert$ or 
$\cM=\vert \cZ_2\vert$, respectively.
Furthermore, when $\cZ_1=\cZ_2=\cZ_3$, the amount of susy preserved is $3/4$ 
and the mass is given by $\cM=\vert\cZ_1\vert$.

Thus, any intersection of branes described in table II can give rise to a 
exotic brane configuration just by considering as the third ``free'' brane 
the one obtained from the product of the initial single projectors. Among 
this set of configurations one can find
\bea
\ba{cccccccccccl}
M5: &1&2&3&4&5&\_&\_&\_&\_&\_ & \nn \\
Mkk: &\_&\_&3&4&5&6&7&8&\_&\_ & \nn \\
M5: &\_&\_&3&4&5&\_&\_&\_&9&\sharp & \,,
\ea
\eea
or
\bea
\ba{cccccccccccl}
Mkk: &1&2&3&4&5&6&\_&\_&\_&\_ & \nn \\
Mkk: &\_&\_&\_&\_&5&6&7&8&9&\sharp & \nn \\
M2: &\_&\_&\_&\_&5&6&\_&\_&\_&\_ & \,.
\ea
\eea

There will generically be an enhancement of supersymmetry by fine tuning
the charges $\{\cZ\}$ whenever some of the commuting single projectors
involved in the configuration are dependent. Such enhancements may include
new exotic fractions of supersymmetry. To illustrate this point, we will
prove that such an enhancement is not possible whenever we have a set
of $n$ independent commuting single projectors $(n\leq 5)$.

As usual, we can simultaneously diagonalize the set of single projectors
giving rise to a set of $2^n$ eigenvalues
\[
\la_i = \pm\cZ_1 \pm \dots \pm \cZ_n \,.
\]
Let $\7\la\geq\la_i$, where $\7\la=\7\ep_1\cZ_1+ \dots + \7\ep_n\cZ_n$.
When the BPS bound is saturated, $\cM=\7\la$ and
\[
\7\la-\la_i = (\7\ep_1-\ep)\cZ_1 + \dots + (\7\ep_n-\ep)\cZ_n\geq 0\,.
\]
By hypothesis, $\cZ_i=\7\ep_i\vert\cZ_i\vert$, so that
\[
\7\la-\la_i = \sum_{i=1}^n \left(1-\7\ep_i\ep_i\right)\vert\cZ_i\vert\geq 0\,.
\]
Whenever signs coincide $(\7\ep_i=\ep_i)$, there is no contribution
to the left hand side summation. Whenever they do not coincide,
$\7\ep_i\ep_i=-1$ so we are left with a sum of positively defined terms
$(2\vert\cZ_i\vert)$. Thus the only possibility to satisfy the equality
in the previous equation is to switch off the charges $\vert\cZ_i\vert=0$,
which corresponds to not having these branes in the configuration. We 
conclude that it is not possible to enhance supersymmetry by fine tuning
of charges whenever the set of $n$ commuting single projectors are
independent, as we claimed. It is straightforward to show that the latter
conclusion does not apply when some of the single projectors are dependent,
as explicitly proved at the beginning of this subsection in a particular
case.

\subsection{Non-factorizable states}

In this subsection, we shall provide an example for a non-factorizable BPS
state. As it has already been stressed in section II, one could have
studied the classification of BPS states in terms of the number of single 
branes building up the state $(N)$ and the number of commutation relations
among the single projectors involved in (\ref{eigen}). The easiest
non-factorizable system that one faces is found for $N=4$ and it is described
by the commutation relations :
\bea
&[\Gamma_1,\Gamma_2]=[\Gamma_1,\Gamma_3]=[\Gamma_3,\Gamma_4]=0 \,,& 
\nonumber \\
&\{\Gamma_1,\Gamma_4\}=\{\Gamma_2,\Gamma_3\}=\{\Gamma_2,\Gamma_4\}=0 &
\label{com}
\eea

One may proceed as in previous subsections to solve the eigenvalue
problem (\ref{eigen}), and one would derive that such state $\state$ exists
and satisfies
\bea
\cP_1\state &=& \frac{ \cZ_1 \Gamma_1 + \cZ_2 \Gamma_2 +  \cZ_3 \Gamma_3 
+  \cZ_4 \Gamma_4}{\cM} \state \nonumber \\
&=& \state \label{proj1} \\
\cP_2\state &=& \frac{2 \cZ_1 \cZ_2 \Gamma_1 \Gamma_2 + 2 \cZ_1 \cZ_3 
\Gamma_1 \Gamma_3 + 2 \cZ_3 \cZ_4 \Gamma_3 \Gamma_4}{\cM^\prime}\state
\nonumber \\
&=& \state \,, 
\label{proj2}
\eea
whereas its mass is given by
\bea
\cM &=& \sqrt{\cZ^{2}_1 + \cZ^{2}_2 + \cZ^{2}_3 + \cZ^{2}_4 + \cM^\prime} \\
\cM^\prime &=& \sqrt{(2 \cZ_1 \cZ_2)^{2} + (2 \cZ_1 \cZ_3)^{2}  + 
(2 \cZ_3 \cZ_4)^{2}}\,.
\eea

At this stage, one is left to determine the amount of supersymmetry
preserved by $\state$. In this case, the analysis is not so straightforward
as in the factorizable case because even though $\cP_2^2=\bI$ and $\mbox{tr}
\cP_2=0$, the same does not hold for $\cP_1$. Its trace is still vanishing,
but
\be
\cP_1^2=a\bI + b\cP_2\,,
\label{cond1}
\ee
where $a=\sum_{i=1}^4 \cZ_i^2/\cM^2$ and $b=\cM^\prime/\cM^2$, so that
$a+b=1$. It is nevertheless still true that $[\cP_1,\,\cP_2]=0$, so that
they can still be simultaneously diagonalized. Take
\[
\cP_2=\mbox{diag}(\underbrace{1,\dots ,1}_{16},
\underbrace{-1,\dots ,-1}_{16})\,.
\]
The latter fixes $\cP_1^2$ through (\ref{cond1}), and since $\cP_1$ is also
diagonal in this basis, $\cP_1$ is determined modulo signs. Actually, $\cP_1$
depends on the number of plus signs appearing in the first and second
$16\times 16$ subspaces $(N_1,\,N_2)$ and also in $a$ or $b$ parameters.
The constraint provided by the vanishing of the trace gives
\be
N_1 + N_2\sqrt{1-2b} = 8(1+\sqrt{1-2b})\,.
\label{cond2}
\ee
Whenever $0\leq b \leq \frac{1}{2}$, but otherwise arbitrary, the solution
to equation (\ref{cond2}) involves $N_1=8$ \footnote{The purpose of this 
subsection is to illustrate the existence of solutions to the eigenvalue
problem (\ref{eigen}) corresponding to non-decoupabble states, and not to
give a full analysis to the projection conditions (\ref{proj1}) and 
(\ref{proj2}).}. Altogether, we conclude that there exist $\nu=\frac{1}{4}$
non-decoupabble BPS states, which were certainly not included in our previous
classification. A very simple configuration satisfying the commutation
relations (\ref{com}) is
\bea
\ba{cccccccccccl}
M2: &1&2&\_&\_&\_&\_&\_&\_&\_&\_ & \nn \\
M2: &\_&2&3&\_&\_&\_&\_&\_&\_&\_ & \nn \\
M2: &\_&\_&3&4&\_&\_&\_&\_&\_&\_ & \nn \\
M2: &\_&\_&\_&4&5&\_&\_&\_&\_&\_ & \,.
\ea
\eea

\section{Automorphisms}

It has already been pointed out \cite{west1,west2,jeromepaul} that the 
maximal automorphism group of the ${\cal N}=1$ $D=11$ SuperPoincar\'e algebra 
is $GL(32,\bR)$. Let us write the forementioned algebra as
\[
\{Q_\alpha,Q_\beta\}=\cZ_{\alpha\beta} \quad ,
\quad [Q_\gamma,\cZ_{\alpha\beta}]=[\cZ_{\alpha\beta},\cZ_{\gamma\delta}]
=0\,.
\]
If we consider the most general transformation on the supercharges,
$Q^\prime_\alpha = (U\,Q)_\alpha$, $U\in \mbox{GL}(32,\bR)$, the latter
will indeed be an automorphism of the algebra if
\begin{equation}
\cZ_{\alpha\beta}^\prime = \left(U\cZ U^t\right)_{\alpha\beta}\,.
\label{trans}
\end{equation}
As stressed in \cite{west1}, any generator of the $GL(32,\bR)$ transformation
can be expanded in terms of the elements of the enveloping Clifford algebra.
This gives rise to $2^{10}$ independent generators, matching the dimension
of $GL(32,\bR)$. In particular, one may consider elements of the subgroup
$SL(32,\bR)$ of the form $U=e^{\alpha\G/2}$ where $\G=\G_{i_1\dots i_p}$
$i_j=1,\dots ,9,\sharp$ $j\leq 1,\dots ,p$ and $p\leq \sharp$. They are indeed
elements of $SL(32,\bR)$ since $\mbox{det}\,U =1$ \footnote{Notice that we only
allowed spacelike indexes in the antisymmetrized gamma matrices defining
the group element $U$, so that $\mbox{tr}\,\G=0$, which guarantees
$\mbox{det}\,U = e^{\mbox{tr}\log U} = e^{\alpha\mbox{tr}\,\G/2}=1$.}. On the 
other hand, there are $2^{10}-1$ independent elements of this type since
\[
\sum_{i=1}^{10} 
\left(\begin{array}{cl}
10 \\
i
\end{array}\right) = 2^{10}-1\,,
\]
the missing element being the identity. It will be important for the rest
of the discussion to distinguish among those $\G$'s being symmetric
$\G_s$ $(p=1,4,5,8,9)$ and those being antisymmetric $\G_a$ $(p=2,3,6,7,10)$.
It is clear that $\G_s^2=1$, whereas $\G_a^2=-1$, so that we can write
\bea
U_s &=& e^{\alpha\G_s/2}= \cosh\frac{\alpha}{2} + \sinh\frac{\alpha}{2}\,
\G_s \\
U_a &=& e^{\alpha\G_a/2}= \cos\frac{\alpha}{2} + \sin\frac{\alpha}{2}\,\G_a\,.
\eea
Furthermore, the subset of elements $U_a$ belongs to an $SO(32)$ subgroup
which leaves the mass operator invariant. That they belong to $SO(32)$ is
clear because $U^t_a = U^{-1}_a$ and that they leave the mass operator
invariant is also clear since
\begin{equation}
U_a\G U^t_a=\G
\end{equation}
whenever $[\G,\G_a]=0$, which is the case for the mass operator $-\cM
\delta_{\alpha\beta}$.

Using this formulation is particularly useful to show that $U$ transformations
leave indeed the \emph{superalgebra covariant}. Notice that 
$\cZ_{\alpha\beta}$ is
expanded in terms of symmetric matrices belonging to the enveloping Clifford 
algebra $\left(\G_s\right)$. We already know that these matrices are left
invariant under $U_a$ if $[\G_s,\G_a]=0$. When $\{\G_s,\G_a\}=0$, then
\begin{equation}
U_a\G_s U^t_a=\cos\alpha\,\G_s + \sin\alpha\,\G_a\G_s\,.
\label{key}
\end{equation}
Due to the fact that $\left(\G_a\G_s\right)^2=\bI$ and 
$\left(\G_a\G_s\right)^t=\G_a\G_s$, one can always write
$\G_a\G_s$ as a symmetric matrix $\G_a\G_s=\G^\prime_s$.

On the other hand, $\G_s$ is left invariant under $U_s=e^{\alpha\G^\prime_s/2}$
whenever $\{\G_s,\G^\prime_s\}=0$. If $[\G_s,\G^\prime_s]=0$, then
\begin{equation}
U_s\G_s U^t_s=\cosh\alpha\,\G_s +\sinh\alpha\,\G_s
\G^\prime_s\,.
\end{equation}
Again $\G_s\G^\prime_s=\G^{\prime\prime}_s$ is a symmetric matrix. Notice,
in particular, that the mass operator transforms under $U_s$ transformations.

Thus $U\cZ U^t$ is a combination of symmetric matrices
which upon requiring the matching with $\cZ^\prime_{\alpha\beta}$ will induce
some transformations on the ``central charges'' $\cZ_{M\dots}$. For example,
let us consider $U_a\in SO(32)$ and its effect on
$\cZ_s\G_s +\cZ^\prime_s\G^\prime_s$, where $\G^\prime_s=\G_s\G_a$ and
$\{\G_s,\G_a\}=0$. An
straightforward computation shows that
\begin{eqnarray}
U_a\left(\cZ_s\G_s +\cZ^\prime_s\G^\prime_s\right)U^t_a &=&
\left(\cZ_s\cos\alpha + \cZ^\prime_s\sin\alpha\right)\G_s \nonumber \\
& + & \left(-\cZ_s\sin\alpha + \cZ^\prime_s\cos\alpha\right)\G^\prime_s\,.
\end{eqnarray}
The latter induces an $SO(2)$ transformation on the space of charges
expanded by $\cZ_s$ and $\cZ^\prime_s$. Indeed,
\begin{equation}
\left(\begin{array}{cl}
\7\cZ_s\\
\7\cZ^\prime_s
\end{array}\right) = R
\left(\begin{array}{cl}
\cZ_s\\
\cZ^\prime_s
\end{array}\right) 
\end{equation}
where
\begin{equation}
R= \left(\begin{array}{ccl}
\cos\alpha & \sin\alpha \\
-\sin\alpha & \cos\alpha
\end{array}\right)\in \mbox{SO}(2)\,.
\end{equation}

Analogously, we could have considered $U_s=e^{\alpha\G^{\prime\prime}_s/2}$
and its effect on $\cZ_s\G_s +\cZ^\prime_s\G^\prime_s$, where 
$\G^\prime_s=\G_s\G^{\prime\prime}_s$ and $[\G_s,\G^{\prime\prime}_2]=0$. 
As before,
\begin{eqnarray}
& U_s\left(\cZ_s\G_s +\cZ^\prime_s\G^\prime_s\right)U^t_s =
\left(\cZ_s\cosh\alpha + \cZ^\prime_s\sinh\alpha\right)\G_s & \nonumber \\
& + \left(\cZ_s\sinh\alpha + \cZ^\prime_s\cosh\alpha\right)\G^\prime_s \,, &
\end{eqnarray}
which in this case involves an $SO(1,1)$ transformation on the space of charges
expanded by $\cZ_s$ and $\cZ^\prime_s$. Indeed,
\begin{equation}
\left(\begin{array}{cl}
\7\cZ_s\\
\7\cZ^\prime_s
\end{array}\right) = S
\left(\begin{array}{cl}
\cZ_s\\
\cZ^\prime_s
\end{array}\right) 
\end{equation}
where
\begin{equation}
S= \left(\begin{array}{ccl}
\cosh\alpha & \sinh\alpha \\
\sinh\alpha & \cosh\alpha
\end{array}\right)\in \mbox{SO}(1,1)\,.
\end{equation}

Let us reexamine eq. (\ref{key}) when $\G_a=\G_s^\prime\G_s$. Since
$\{\G_a,\G_s\}=0$ by hypothesis, $\{\G_s,\G_s^\prime\}=0$ and eq.
(\ref{key}) looks like
\begin{equation}
U_a\G_s U^t_a = \cos\alpha\,\G_s + \sin\alpha\,\G^\prime_s\,,
\label{key1}
\end{equation}
this being nothing else but a linear combination of two anticommuting
single projectors $(\G_s,\G_s^\prime)$ with coefficients parametrizing
a circle. The latter was the operator $\cP$ describing a non-threshold
bound state $\state$. Since $\state$ is a Clifford
valued state, we can look at $\cP\state=\state$ as
\begin{equation}
U_a\G_s U^t_a\state = \state \Leftrightarrow \G_s\left(U^t_a\state\right)
=\left(U^t_a\state\right)\,,
\end{equation}
which tells us that the initial $\nu=\frac{1}{2}$ non-threshold bound state
$\state$ is $SO(32)$ related with a  $\nu=\frac{1}{2}$ single brane (at
threshold) $\vert\alpha^\prime>=U^t_a\state$ having the same mass.

As a first consequence, we can immediately state that any non-threshold
bound state appearing in table \ref{table3} is $SO(32)$ related with a 
$\nu=\frac{1}{2}$ bound state at threshold. Let us analyze some particular
examples of this phenomena. Consider the configuration
\bea
\ba{cccccccccccl}
M2: &1&2&\_&\_&\_&\_&\_&\_&\_&\_ & \nn \\
M2: &\_&2&3&\_&\_&\_&\_&\_&\_&\_ & 
\ea
\eea
characterized by $\cP=\cos\alpha\,\G_{012} + \sin\alpha\,\G_{023}$
and $\cM=\sqrt{\cZ_{12}^2 +\cZ_{23}^2}$. It is $SO(32)$ related
with a single M2-brane of the same mass through $U_\alpha=e^{\alpha\G_{13}/2}$,
which corresponds to a rotation of angle $\alpha=\mbox{arccos}\left(
\frac{\cZ_{12}}{\cM}\right)$ in the $13$-plane.

There exist much more exotic transformations such as the one relating
\bea
\ba{cccccccccccl}
M5: &1&2&3&4&5&\_&\_&\_&\_&\_ & \nn \\
M2: &1&2&\_&\_&\_&\_&\_&\_&\_&\_ & 
\ea
\eea 
to a single M5-brane through $U_\alpha=e^{\alpha\G_{345}/2}$, which is
reminiscent of the electro-magnetic duality transformation in eight dimensions
(dyonic membranes) \cite{lpaul} or the one relating
\bea
\ba{cccccccccccl}
Mkk: &1&2&3&4&5&6&\_&\_&\_&\_ & \nn \\
M2:  &\_&\_&\_&\_&\_&6&7&\_&\_&\_ & 
\ea
\eea 
to a single Mkk-monopole through $U_\alpha=e^{\alpha\G_{123457}/2}$.

Actually, this construction can be extended to a linear combination
of an arbitrary number $(n+1)$ of mutually anticommuting single projectors,
with coefficients parametrizing $S_n$. Indeed, consider
\bea
\cP&=&\cos\alpha_1\,\G_1 + \sin\alpha_1\left(\cos\alpha_2\,\G_2 +
\sin\alpha_2\left(\dots + \right.\right. \nonumber \\
& & \left.\left.\sin\alpha_{n-1}\left(\cos\alpha_n\,\G_n +
\sin\alpha_n\,\G_{n+1}\right)\dots\right)\right)\,. \nonumber
\eea
Using identity (\ref{key1}) iteratively,
\bea
\cos\alpha_n\,\G_n + \sin\alpha_n\,\G_{n+1}&=&U_{\alpha_n}\,\G_n\,, 
\nonumber \\
U_{\alpha_n}&=&e^{\alpha_n\G_{n+1}\G_{n}} \\
\cos\alpha_{n-1}\,\G_{n-1} + \sin\alpha_{n-1}\,U_{\alpha_n}\G_n &=&
U_{\alpha_{n-1}}\,\G_{n-1}\,, \nonumber \\
U_{\alpha_{n-1}}&=&e^{\alpha_{n-1}U_{\alpha_n}\G_{n}\G_{n-1}} \\
\quad . & & \quad . \nonumber \\
\quad . & & \quad . \nonumber \\
\quad . & & \quad . \nonumber \\
\cos\alpha_1\,\G_1 + \sin\alpha_1\,U_{\alpha_2}\G_2 &=&
U_{\alpha_1}\,\G_1\,, \nonumber \\
U_{\alpha_1}&=&e^{\alpha_1 U_{\alpha_2}\G_2\G_1}\,,
\eea
we can rewrite $\cP$ as
\be
\cP= U_{\alpha_1}\G_1 = U_{\alpha_1/2}\,\G_1\,
U^t_{\alpha_1/2}\,.
\ee
Thus any non-threshold bound state preserving $\nu=\frac{1}{2}$ characterized
by $\cP\state=\state$ is $SO(32)$ related with a $\nu=\frac{1}{2}$ bound
state at threshold having the same mass.

We shall now move to less supersymmetric BPS states and study whether we can
find more involved $\mbox{SO}(32)$ transformations relating non-threshold
bound states to bound states at threshold. Let us start by BPS states
described by (\ref{m42}) and (\ref{m4b}). Without loss of generality, we
shall concentrate on operators $\cP_i$ collapsing to single projectors
$\left(\cP_i=\G_i\right)$
\bea
\G_3\G_4\state&=&\state \nonumber \\
\left(\cos\8\alpha\,\G_1 + \sin\8\alpha\,\G_3\right)\state&=&\state \,,
\eea
where $\tan\8\alpha=\frac{\cZ_3 + \cZ_4}{\cZ_1}$ and whose mass is given
by
\be
\cM= \sqrt{\left(\cZ_1\right)^2 + \left(\cZ_3 + \cZ_4\right)^2} \,.
\ee

Let us compute the transformed charges $\left(\cZ^\prime\right)$ under the
finite $\mbox{SO}(32)$ transformation
\[
U=e^{\alpha\G_3\G_1/2}e^{\beta\G_4\G_1/2}
\]
for arbitrary $\alpha$, $\beta$ parameters. Using (\ref{trans}), we obtain
\bea
\cZ_1^\prime &=& \cZ_1\cos\alpha\cos\beta - \cZ_3\sin\alpha\cos\beta-
\cZ_4\cos\alpha\sin\beta \nonumber \\
\cZ_3^\prime &=& \cZ_1\sin\alpha\cos\beta + \cZ_3\cos\alpha\cos\beta-
\cZ_4\sin\alpha\sin\beta \nonumber \\
\cZ_4^\prime &=& \cZ_1\cos\alpha\sin\beta - \cZ_3\sin\alpha\sin\beta+
\cZ_4\cos\alpha\cos\beta \nonumber \\
\cZ_{134}^\prime &=& -\cZ_1\sin\alpha\sin\beta - \cZ_3\cos\alpha\sin\beta-
\cZ_4\sin\alpha\cos\beta \, , \nonumber \\
\label{trans1}
\eea
where $\cZ_{134}^\prime$ is the central charge associated with the single
projector $\G_1\G_3\G_4$. We are thus led to four independent charges.
Actually, we can appropiately fix $\alpha$ and $\beta$ to set two of the
$\cZ^\prime$'s to zero. In particular, from the requirements $\cZ_3^\prime +
\cZ_4^\prime=0$ and $\cZ_3^\prime - \cZ_4^\prime=0$, we derive two
conditions
\bea
\tan\left(\alpha+\beta\right)&=&-\frac{\cZ_3+\cZ_4}{\cZ_1}
\label{e1} \\
\tan\left(\alpha-\beta\right)&=&\frac{\cZ_4-\cZ_3}{\cZ_1}\,,
\label{e2}
\eea
fixing both arbitrary parameters. Notice that $\8\alpha=-
\left(\alpha+\beta\right)$. It can be checked that the invariant mass $\cM$, 
when expressed in terms of the transformed charges, looks as
\[
\cM=\cZ_1^\prime + \cZ_{134}^\prime\,,
\]
which is reminiscent of a $\nu=\frac{1}{4}$ BPS state at threshold. This
expectation can be further confirmed by analysing the projection conditions
satisfied by $\vert\alpha^\prime>=U^t\state$. Since $[\G_3\G_4,U]=0$,
\be
\G_3\G_4\vert\alpha^\prime>=\vert\alpha^\prime>\,.
\label{e11}
\ee
On the other hand,
\bea
& U^t\left(\cos\8\alpha\G_1+\sin\8\alpha\G_3\right)U= & \nonumber \\
& \cos(\8\alpha+\alpha+\beta)\G_1+\sin(\8\alpha+\alpha+\beta)\G_3=\G_1 &
\eea
thus,
\bea
U^t\left(\cos\8\alpha\G_1+\sin\8\alpha\G_3\right)U\vert\alpha^\prime>
&=&\vert\alpha^\prime> \nonumber \\
& \Updownarrow & \nonumber \\
\G_1\vert\alpha^\prime>&=&
\vert\alpha^\prime>\,.
\label{e12}
\eea
Joining equations (\ref{e11}) and (\ref{e12}), the transformed BPS
state is described by
\bea
\G_1\vert\alpha^\prime>&=&\vert\alpha^\prime> \nonumber \\
\G_1\G_3\G_4\vert\alpha^\prime>&=&\vert\alpha^\prime> \nonumber \\
\cM &=&\cZ_1^\prime + \cZ_{134}^\prime\,,
\eea
corresponding to a $\nu=\frac{1}{4}$ BPS state at threshold.

Even though the latter proof applies to any set of $\{\G_1,\,\G_3,\,\G_4\}$
satisfying $\{\G_1,\,\G_3\}=\{\G_1,\,\G_4\}=[\G_3,\,\G_4]=0$, it may be
useful to consider some particular examples. One involving just 
$\mbox{SO}(10)$ rotations is
\bea
\ba{cccccccccccl}
M2: &\_&2&3&\_&\_&\_&\_&\_&\_&\_ & \nn \\
M2: &1&2&\_&\_&\_&\_&\_&\_&\_&\_ & \nn \\
M2: &\_&\_&3&4&\_&\_&\_&\_&\_&\_ & \,.
\ea
\eea
This non-threshold $\nu=\frac{1}{4}$ BPS state is related by the
transformation
\bea
U&=&e^{-\alpha\G_{13}/2}e^{\beta\G_{24}/2} \nonumber \\
\tan\left(\alpha+\beta\right) &=& -\frac{\cZ_{12}+\cZ_{34}}{\cZ_{23}}
\nonumber \\
\tan\left(\alpha-\beta\right) &=& \frac{\cZ_{34}-\cZ_{12}}{\cZ_{23}}\,,
\eea
to the BPS state at threshold described by
\bea
\G_{023}\vert\alpha^\prime>&=&\vert\alpha^\prime> \nonumber \\
\G_{014}\vert\alpha^\prime>&=&\vert\alpha^\prime> \nonumber \\
\cM &=& \cZ_{23}^\prime + \cZ_{14}^\prime
\eea
corresponding to the array $\mbox{M2}\perp\mbox{M2}(0)$
\bea
\ba{cccccccccccl}
M2: &\_&2&3&\_&\_&\_&\_&\_&\_&\_ & \nn \\
M2: &1&\_&\_&4&\_&\_&\_&\_&\_&\_ & \,.
\ea
\eea
Another less trivial example is provided by the array
\bea
\ba{cccccccccccl}
M2: &1&2&\_&\_&\_&\_&\_&\_&\_&\_ & \nn \\
M5: &1&\_&3&4&5&6&\_&\_&\_&\_ & \nn \\
M2: &1&\_&3&\_&\_&\_&\_&\_&\_&\_ & \,.
\ea
\eea
The latter can be $\mbox{SO}(32)$ related to
\bea
\ba{cccccccccccl}
M2: &1&\_&3&\_&\_&\_&\_&\_&\_&\_ & \nn \\
M5: &1&2&\_&4&5&6&\_&\_&\_&\_ & \,,
\ea
\eea
through $U=e^{\alpha\G_{23}/2}e^{-\beta\G_{456}/2}$ with $\alpha$, $\beta$
satisfying
\bea
\tan\left(\alpha+\beta\right) &=& -\frac{\cZ_{12}+\cZ_{13456}}{\cZ_{13}}
\nonumber \\
\tan\left(\alpha-\beta\right) &=& \frac{\cZ_{13456}-\cZ_{12}}{\cZ_{13}}\,.
\eea

The above construction can be extended to non-threshold $\nu=\frac{1}{8}$
BPS states described by
\bea
\G_1\G_2\state &=&\state \nonumber \\
\G_3\G_4\state &=&\state \nonumber \\
\left(\cos\8\alpha\,\G_1+\sin\8\alpha\,\G_3\right)\state \,,
&=&\state 
\eea
whose mass is given by
\be
\cM = \sqrt{\left(\cZ_1+\cZ_2\right)^2 +
\left(\cZ_3+\cZ_4\right)^2}\,.
\ee
Proceeding as in the previous case, one can compute the transformed
charges $\left(\cZ^\prime\right)$ under the finite $\mbox{SO}(32)$
transformation
\[
U=e^{\alpha\G_3\G_1/2}e^{\beta\G_4\G_1/2}e^{\gamma\G_3\G_2/2}
e^{\delta\G_4\G_2/2}
\]
for arbitrary $\alpha$, $\beta$, $\gamma$, $\delta$ parameters. The number
of independent transformed charges $\left(\cZ^\prime\right)$ is eight. The
new four ones $\{\cZ^\prime_{123},\,\cZ^\prime_{124},\,\cZ^\prime_{134},
\,\cZ^\prime_{234}\}$ are associated with the single projectors
$\{\G_1\G_2\G_3,\,\G_1\G_2\G_4,\,\G_1\G_3\G_4,\,\G_2\G_3\G_4\}$ respectively.
Requiring
\[
\cZ^\prime_3 + \ep_1\cZ^\prime_4 + \cZ^\prime_{123} + \ep_2\cZ^\prime_{124}
=0
\]
$\forall$ $\ep_1,\,\ep_2$ satisfying $\ep_1^2=\ep_2^2=1$, one can fix the four
parameters
\[
\tan[\alpha+\ep_1\beta+\ep_2(\gamma+\ep_1\delta)]=-
\frac{\cZ_3+\ep_1\cZ_4}{\cZ_1+\ep_2\cZ_2}\,.
\]
At this stage, one is left with four non-vanishing charges 
$\{\cZ^\prime_{1},\,\cZ^\prime_{2},\,\cZ^\prime_{134},\,\cZ^\prime_{234}\}$
associated with the four commuting single projectors
$\{\G_1,\,\G_2,\,\G_1\G_3\G_4,\,\G_2\G_3\G_4\}$. When reexpressing the
mass in terms of these charges, one derives
\[
\cM=\cZ^\prime_1+\cZ^\prime_2+\cZ^\prime_{134}+\cZ^\prime_{234}\,.
\]
All the above information seems to indicate that the transformed state
preserves $\nu=\frac{1}{16}$, but this conclusion is incorrect. If one
analyses the transformed supersymmetry projection conditions satisfied by
$\vert\alpha^\prime>=U^t\state$ one is left with
\be
\begin{array}{cccl}
\G_1\vert\alpha^\prime>=\vert\alpha^\prime> & & 
\G_1\vert\alpha^\prime>=\vert\alpha^\prime> \nonumber \\
\G_1\G_2\vert\alpha^\prime>=\vert\alpha^\prime> &
\Leftrightarrow & \G_2\vert\alpha^\prime>=\vert\alpha^\prime> \nonumber \\
\G_3\G_4\vert\alpha^\prime>=\vert\alpha^\prime> & &
\G_1\G_3\G_4\vert\alpha^\prime>=\vert\alpha^\prime>  
\end{array}
\ee
from which we can appreciate that 
$\G_2\G_3\G_4\vert\alpha^\prime>=\vert\alpha^\prime>$ is an extra projection
condition that can be added for free, thus not breaking further
supersymmetry. This proves our claim concerning this particular non-threshold
$\nu=\frac{1}{8}$ BPS state. There should be other $\mbox{SO}(32)$ 
transformations, responsible for the same phenomena, for more involved
and less supersymmetric non-threshold bound states.

We would like to finish with a brief remark concerning factorizable
states at threshold. Assume their mutually commuting subsystems
are $\mbox{SO}(32)$ related with subsystems at threshold through
$U_i\in \mbox{SO}(32)$ transformations, then the full system is
$\mbox{SO}(32)$ related with a set of commuting single branes
through the automorphism 
\[
U=\prod_{i=1}^{n}U_i \quad (n\leq 5)\,.
\]
The essential ingredient of the proof is to consider two commuting
projectors $\cP_1\,,\cP_2$ such that
\bea
\cP_1 &=& U_1\,\G_1\,U_1^t \nonumber \\
\cP_2 &=& U_2\,\G_2\,U_2^t\,,
\eea
where $\G_1^2=\G_2^2=\bI$, $\mbox{tr}\,\G_1 =\mbox{tr}\,\G_2 = 0$. By 
hypothesis, not only $[\cP_1\,,\cP_2]$ vanishes but also any commutator
of any single projectors in $\cP_1$ with any single projector in $\cP_2$.
Under these circumstances, $[\G_1\,,U_2]=[\cP_2\,,U_1]=0$ which are sufficient
conditions to prove that the overall $SO(32)$ automorphism relating the
state $\cP_1\state=\cP_2\state=\state$ to
\[
\G_1\vert\alpha^\prime>=\G_2\vert\alpha^\prime>=\vert\alpha^\prime>
\]
where $\state=U\,\vert\alpha^\prime>$ with $U=U_1\,U_2$.
The extension to a set of $n$ $(n\leq 5)$
commuting projectors $\{\cP_i\}$ is straightforward.

\section{Discussion}

As happens with R-symmetry in supersymmetric field theories \cite{weinberg3},
the automorphism group of a given superalgebra may or may not be a good 
symmetry of the theory. If it is, it may be violated by anomalies or
spontaneously broken, or it may remain a good symmetry of the theory. It
remains an open question to know whether the group $GL(32,\bR)$, or a
subgroup of it, is a symmetry of M-theory and if so, which formulation would 
make it manifest. In the following, we shall not try to answer these
questions but we shall point out some remarks concerning the possible
realization of automorphisms on world volume effective actions.

It is well-known that kappa invariant world volume brane theories do provide
us with field theory realizations of the corresponding spacetime supersymmetry
algebras. BPS states are realized in brane theory through field configurations
solving the kappa symmetry preserving condition \cite{renatatomas}
\[
\G_\kappa\ep=\ep\, ,
\]
and saturating the bound on the energy \cite{quimjeromepaul}.

Thus, branes propagating in ${\cal N}=1$ $D=11$ SuperPoincar\'e give us
field theory realizations of the corresponding SuperPoincar\'e algebra, or
truncations of it. It is then natural to wonder whether the group of
automorphisms of such an algebra is a symmetry of the corresponding
world volume theory. Since the Lorentz group in eleven dimensions can be
seen as a subgroup of $\mbox{GL}(32,\bR)$, it is obvious that such
subgroup will be linearly realized on the brane (before any gauge fixing).

In the previous section, we showed that central charges $\cZ$'s are
generically ``rotated'' among themselves under automorphism transformations.
Since for bosonic configurations, such topological charges are given by
world space integrals involving derivatives of the brane dynamical fields,
one should also expect, if any, the existence of non-local transformations
leaving certain brane theories invariant. This is the case for the non-local
transformations leaving the D3-brane effective action invariant 
\cite{kiyoshi}. The latter are the world volume realization of the S-duality
automorphism for the ${\cal N}=2$ $D=10$ IIB SuperPoincar\'e algebra. The
analysis performed in \cite{joan} shows that there should be
similar non-local transformations giving rise to symmetry transformations
of other D-brane effective actions by T-duality. We hope to come to these
issues in the future.

%%%%%%%%%%%%%%%%%%%%%%%%%%%%%%%%%%%%%%%%%%%%%%%%%%%%%%%

\emph{Acknowledgements}

JM and JS are supported by a fellowship from Comissionat per a Universitats i 
Recerca de la Generalitat de Catalunya. This work was supported in part by
AEN98-0431 (CICYT), GC 1998SGR (CIRIT).

%%%%%%%%%%%%%%%%%%%%%%%%%%%%%%%%%%%%%%%%%%%%%%%%%%%%%%%%%%%%%%%%%%

\begin{table}
\begin{tabular}{|c|c|c|}
Brane & Projector & Charge \\
\hline
Mw & $\G_{0m_1}$ & $\cZ^{m_1}$ \\
\hline
M2 & $\G_{0m_1m_2}$ & $\cZ^{m_1m_2}$ \\
\hline
M5 & $\G_{0m_1\dots m_5}$ & $\cZ^{m_1\dots m_5}$ \\
\hline
Mkk & $\G_{0m_1\dots m_6}$  & $\cZ^{0m_7\dots m_\sharp}$ \\
\hline
M9 & $\G_{0m_1\dots m_9}$ & $\cZ^{0m_\sharp}$ \\
\end{tabular}
\medskip
\caption{Single branes preserving $\nu=\frac{1}{2}$, their supersymmetry
projection condition and their charges.}
\label{table1}
\end{table}

\begin{table}
\begin{tabular}{|c|c|c|c|c|}
$\perp$ & M2 & M5 & Mkk & M9 \\
\hline
Mw & 1 & 1 & 1 & 1 \\
\hline
M2 & 0 & 1 & 0,2 & 1  \\
\hline
M5 & 1 & 1,3 & 1,3,5 & 5 \\
\hline
Mkk & 0,2 & 1,3,5 & 2,4 & 5 \\
\end{tabular}
\medskip
\caption{Threshold bound states involving two single branes.}
\label{table2}
\end{table}

\begin{table}
\begin{tabular}{|c|c|c|c|c|c|}
$\perp$ & Mw & M2 & M5 & Mkk & M9 \\
\hline
Mw & 0 & 0 & 0& 0 & 0 \\
\hline
M2 & 0 & 1 & 0,2 & 1 & 2 \\
\hline
M5 & 0 & 0,2 & 0,2,4& 2,4 & 4 \\
\hline
Mkk & 0 & 1 & 2,4 & 3,5 & 6 \\
\hline
M9 & 0 & 2 & 4 & 6 & 8 \\
\end{tabular}
\medskip
\caption{Non-threshold bound states involving two single branes.}
\label{table3}
\end{table}


\begin{thebibliography}{99}
\bibitem{dual} C. Hull and P. K. Townsend, Nucl. Phys. {\bf B438} (1995)
109 (hep-th/9505073);\par
E. Witten, Nucl. Phys. {\bf B443} (1995) 85 (hep-th/9503124).
\bibitem{bhen} A. Strominger and C. Vafa, Phys. Lett. {\bf B379} (1996) 99
(hep-th/9601029);\par
C. Callan and J. M. Maldacena, Nucl. Phys. {\bf B472} (1996) 591 
(hep-th/9602043);\par
J. C. Breckenridge, R. C. Myers, A. W. Peet and C. Vafa, Phys. Lett. {\bf B391}
(1997) 93 (hep-th/9602065);\par
J. M. Maldacena and A. Strominger, Phys. Rev. Lett. {\bf 77} (1996) 428
(hep-th/9603060);\par
C. V. Johnson, R. R. Khuri and R. C. Myers, Phys. Lett. {\bf B378} (1996)
78 (hep-th/9603061).
\bibitem{examp} G. Papadopoulos and P. K. Townsend, Phys. Lett. {\bf B380}
(1996) 273 (hep-th/9603087);\par
A. A. Tseytlin, Nucl. Phys. {\bf B475} (1996) 149 (hep-th/9604035);\par
I. R. Klebanov and A. A. Tseytlin, Nucl. Phys. {\bf B475} (1996) 179 
(hep-th/9604166);\par
V. Balasubramanian and F. Larsen, Nucl. Phys. {\bf B478} (1996) 199
(hep-th/9604189);\par
J. P. Gauntlett, D. A. Kastor and J. Traschen, Nucl. Phys. {\bf B478} (1996)
544 (hep-th/9604179);\par
J. X. Lu and S. Roy, JHEP9908 (1999) 002 (hep-th/9904112); Nucl. Phys.
{\bf B560} (1999) 181 (hep-th/9904129); JHEP0001 (2000) 034 (hep-th/9905014);
Phys. Rev. {\bf D60} (1999) 126002 (hep-th/9905056).
\bibitem{eduardo} E. Bergshoeff, M. de Roo, E. Eyras, B. Janssen and
J. P. van der Schaar, Nucl. Phys. {\bf B494} (1997) 119 (hep-th/9612095);
Class. Quant. Grav. {\bf 14} (1997) 2757 (hep-th/9704120).
\bibitem{tech} J. G. Russo and A. A. Tseytlin, Nucl. Phys. {\bf B490} (1997)
121 (hep-th/9611047);\par
M. S. Costa and G. Papadopoulos, Nucl. Phys. {\bf B510} (1998) 217
(hep-th/9612204);\par
M. S. Costa and M. Cveti$\7c$, Phys. Rev. {\bf D56} (1997) 4834
(hep-th/9703204).
\bibitem{paulalgebra} P. K. Townsend, \emph{M-theory from its
superalgebra}, in 'Strings, branes and dualities', Carg\`ese 1997, ed. L. 
Baulieu et al., Kluwer Academic Publ. 1999, p.141 (hep-th/9712004).
\bibitem{gauntletthull} J. P. Gauntlett and C. M. Hull, JHEP 0001(2000) 004 
(hep-th/9909098);\par
I. Bandos and J. Lukierski, Mod. Phys. Lett. {\bf A14} (1999) 1257.
\bibitem{west1} O. Baerwald and P. West, Phys. Lett. {\bf B476} (2000) 157
(hep-th/9912226).
\bibitem{west2} P. West, \emph{Automorphisms, Non-Linear Realizations
and Branes}, hep-th/0001216;\par
P. West, \emph{Hidden Superconformal symmetry in M theory}, hep-th/0005270.
\bibitem{democracy} P. K. Townsend, \emph{p-brane democracy}, in 
\emph{Particles, Strings and Cosmology}. eds. J. Bagger, G. Domokos, A. Falk
and S. Kovesi-Domokos (World Scientific 1996), pp. 271-285, (hep-th/9507048).
\bibitem{toine} J. W. Holten and A. Van Proeyen, J. Phys. A: Math Gen. {\bf 15}
(1982) 3763.
\bibitem{jeromepaul} J. P. Gauntlett, G. W. Gibbons, C. M. Hull and
P. K. Townsend, \emph{BPS states of $D=4$ $N=1$ supersymmetry}, hep-th/0001024.
\bibitem{pioline} N. A. Obers and B. Pioline, Phys. Rep. 318 (1999) 113
(hep-th/9809039).
\bibitem{angles} M. Berkooz, M. R. Douglas and R. G. Leigh, Nucl. Phys.
{\bf B480} (1996) 265 (hep-th/9606139);\par
N. Ohta and P. K. Townsend, Phys. Lett. {\bf B418} (1998) 77 (hep--th/9710129)
;\par
B. S. Acharya, J. M. Figueroa-O'Farrill and B. Spence, 
JHEP 9804(1998) 012 (hep-th/9803260); JHEP 9807(1998) 004 (hep-th/9805073);\par
B.S. Acharya, J.M. Figueroa-O'Farrill, B.Spence and S. Stanciu,
JHEP 9807(1998) 005 (hep-th/9805176).
\bibitem{lpaul} J.M. Izquierdo, N.D. Lambert, G. Papadopoulos and
P.K. Townsend, Nucl. Phys. {\bf B460} (1996) 560 (hep-th/9508177).
\bibitem{weinberg3} S. Weinberg, \emph{The quantum theory of fields.
Volume III Supersymmetry}, Cambridge University Press (2000).
\bibitem{renatatomas} E. Bergshoeff, R. Kallosh, T. Ort\'{\i}n and 
G. Papadopoulos, Nucl. Phys. {\bf B502} (1997) 149 (hep-th/9705040).
\bibitem{quimjeromepaul} J. Gauntlett, J. Gomis and P. K. Townsend,
JHEP9801 (1998) 003 (hep-th/9711205).
\bibitem{kiyoshi} Y. Igarashi, K. Itoh and K. Kamimura, Nucl. Phys. 
{\bf B536} (1998) 469 (hep-th/9806161).
\bibitem{joan} J. Sim\'on, Phys. Rev. {\bf D61} 047702 (2000) 
(hep-th/9812095) ;\par
K. Kamimura and J. Sim\'on, Nucl. Phys. {\bf B585} (2000) 213 
(hep-th/0003211).
\end{thebibliography}
\end{document}